\documentclass[apj]{emulateapj}

\slugcomment{}

\shorttitle{}
\shortauthors{}

\usepackage{color}
\usepackage{amsmath}
\usepackage{booktabs}

\begin{document}


\title{Properties of the Molecular Cores of Low Luminosity Objects}

\author{Tien-Hao Hsieh$^{1, 2}$, Shih-Ping Lai$^{1}$, Arnaud Belloche$^{2}$, Friedrich Wyrowski$^{2}$, and Chao-Ling Hung$^{3, 4}$}
\affil{$^{1}$Institute of Astronomy, National Tsing Hua University (NTHU), Hsinchu 30013, Taiwan}
\affil{$^{2}$Max-Planck-Institut f$\ddot{\textmd{u}}$r Radioastronomie (MPIfR), Bonn, Germany}
\affil{$^{3}$Institute for Astronomy, University of Hawaii, USA}
\affil{$^{4}$Harvard-Smithsonian Center for Astrophysics, Cambridge, MA 02138, USA}

\email{slai@phys.nthu.edu.tw, shawinchone@gmail.com}

\begin{abstract}
We present a survey toward 16 Low Luminosity Objects (LLOs with an internal luminosity, $L_{\rm int}$, lower than 0.2 $L_\odot$) with N$_2$H$^+$ (1--0), N$_2$H$^+$ (3--2), N$_2$D$^+$ (3--2), HCO$^+$ (3--2) and HCN (3--2) using the Arizona Radio Observatory Kitt Peak 12m Telescope and Submillimeter Telescope.
Our goal is to probe the nature of these faint protostars which are believed to be either very low mass or extremely young protostars.
We find that the N$_2$D$^+$/N$_2$H$^+$ column density ratios of LLOs are similar to those of typical starless cores and Class 0 objects.
The N$_2$D$^+$/N$_2$H$^+$ column density ratios are relatively high ($>$ 0.05) for LLOs with kinetic temperatures less than 10 K in our sample.
The distribution of N$_2$H$^+$ (1--0) line widths spreads between that of starless cores and young Class 0 objects.
If we use the line width as a dynamic evolutionary indicator, LLOs are likely young Class 0 protostellar sources.
We further use the optically thick tracers, HCO$^+$ (3--2) and HCN (3--2), to probe the infall signatures of our targets.
We derive the asymmetry parameters from both lines and estimate the infall velocities by fitting the HCO$^+$ (3--2) spectra with two-layer models.
As a result, we identify eight infall candidates based on the infall velocities and seven candidates have infall signatures supported by asymmetry parameters from at least one of HCO$^+$ (3--2) and HCN (3--2).
\end{abstract}



\section{INTRODUCTION}
Very Low Luminosity Objects (VeLLOs) are embedded protostars with internal luminosity $L_{\rm int}$ $<$ 0.1 $L_{\odot}$ \citep{di07}.
The first VeLLO, L1014-IRS, was identified by \citet{yo04} toward a dense core previously thought to be starless, L1014.
The SED fitting results suggest that L1014-IRS has a very low internal luminosity $\sim$ 0.09 $L_\odot$ contributed by the central star and disk \citep{yo04,hu06}.
Three possibilities are proposed to explained VeLLOs' low luminosities \citep{du14}: they can be 1) very low mass protostars, 2) extremely young protostars, or 3) protostars in a quiescent phase of an episodic accretion process.
\citet{an12} identified a ``{\it pre-brown dwarf}'', Oph-B11, and suggested that brown dwarfs can form in isolated cores as low mass stars \citep{pa04}.
By comparing with evolutionary track from models, the low luminosities imply that VeLLOs are substellar sources (current mass $\leq$ 0.08 $M_\odot$) embedded in molecular cores (L1014-IRS: Young et al.\ 2004; Huard et al.\ 2006, L1521F: Bourke et al.\ 2006).
\citet{le13} suggested that the mass of L328 is at most 0.05 $M_\odot$ based on the estimate of mass loss rate from outflows.
Therefore, VeLLOs could form low mass stars or even brown dwarfs depending on their future accretion.
VeLLOs with small mass accretion reservoir (parent cores) are very likely brown dwarfs still in the embedded phase, {\it i.e. ``proto brown dwarfs}'' (L328: Lee et al. 2009, 2013, L1148: Kauffmann et al. 2011, ICE 348-SMM2E: Palau et al. 2014).
Since VeLLOs are usually isolated, they could be brown dwarfs forming through cloud fragmentation followed by gravitational collapse.

On the other hand, VeLLOs' substellar masses also indicate that VeLLOs can be early Class 0 protostars or even younger than Class 0 protostars (IRAM 04191: Andr\'{e} et al.\ 1999; Belloche et al.\ 2002; Dunham et al.\ 2006, L1521F: Bourke et al.\ 2006; Takahashi et al.\ 2013, Cha-MMS1: Belloche et al.\ 2006; Tsitali et al.\ 2013; V\"{a}is\"{a}l\"{a} et al.\ 2014). 
Considering several age indicators such as $T_{\rm bol}$ and $L_{\rm bol}$, \citet{an99} suggested that IRAM 04191 is a young Class 0 object with an age of $\lesssim$ 3--5 $\times$ 10$^4$ yr.
\citet{be02} later confirmed this by considering the inner free fall region expanding at the sound speed.
By studying the outflows, \citet{ta13} found that L1521F is probably a protostar in the earliest evolutionary stage (age $\lesssim$ 10$^4$ yr).
\citet{be06} found a high level of deuterium fraction and no sign of large scale outflows for the VeLLO Cha-MMS1, suggesting that it 
has not driven any outflow yet and could be younger than IRAM 04191 and L1521F.
The kinematical study in \citet{ts13} further supports that Cha-MMS1 is either a young Class 0 source or even a first-hydrostatic core (FHSC) \citep{la69}. 
The FHSC phase represents the short-lived phase during which a hydrostatic compact object forms at the center of a collapsing dense core but has not yet experienced the dissociation of H$_2$. \citep{la69}. The dissociation of H$_2$ at the end of the FHSC phase triggers the second collapse that leads to the formation of the protostar.
Several FHSC candidates have been identified but none has been firmly confirmed yet (Cha-MMS1: Belloche et al.\ 2006, 2011a, L1448-IRS2E: Chen et al.\ 2010, Per-Bolo 58: Enoch et al.\ 2010; Dunham et al.\ 2011, L1451-mm: Pineda et al.\ 2011, CB17-MMS: Chen et al.\ 2012, B1-bS, and B1-bN: Pezzuto et al.\ 2012, Murillo \& Lai 2013).

The third possibility -- a protostar in a quiescent accretion phase -- is possible because the mass accretion can enhance the luminosity of protostars, {\it i.e.} the accretion luminosity L$_{\rm acc}$ (L673-7: Dunham et al.\ 2010a, CB130: Kim et al.\ 2011).
Episodic accretion models \citep{du10b,du12} have been proposed to interpret the low luminosities of VeLLOs and could be a possible solution of the luminosity problem \citep{ke90}.
The accretion luminosities estimated based on observed CO outflows further support that the accretion process is episodic in low luminosity objects \citep{du06,du10a,le10}.
Note that these three possibilities do not conflict with each other; VeLLOs could be a mixture of these three possibilities \citep{du14}.

The most thorough VeLLO survey has been done by \citet{du08}.
The Spitzer Legacy Project ``From Molecular Cores to Planet Forming Disks'' (c2d; Evans et al. 2003, 2009) mapped five nearby molecular clouds and 95 small dense cores, which provides an excellent opportunity for searching for VeLLOs.
\citet{du08} found a tight relation between the 70 $\mu$m flux of protostars and their internal luminosity $L_{\rm int}$ = 3.3 $\times$ 10$^8$ $F^{0.94}_{70}$ $L_\odot$  
based on 1460 synthetic SEDs from the Monte Carlo dust radiative transfer code RADMC \citep{du04}.
Using this relation, \citet{du08} constructed a list of embedded protostars with low $L_{\rm int}$ and identified 15 VeLLOs from the c2d data.
Lately, \cite{hs13} developed a new YSO identification method to probe faint YSOs and identified seven additional VeLLO candidates from the c2d survey clouds.


In order to understand the nature of VeLLOs, we focus on probing their evolutionary states using chemical and dynamic evolutionary tracers.
It is known that at the center of starless cores, where the temperature is about 10 K and the number density is above 10$^4$ cm$^{-3}$, common molecular tracers such as CO and CS freeze out onto dust grain surfaces \citep{ca99, ta02, ta04}. 
Other tracers are required to investigate the dynamical and chemical properties in the cold dense region.
The nitrogen-bearing species such as N$_2$H$^+$ and NH$_3$ survive in the gas phase at least for densities in the range of 10$^5 -$10$^6$ cm$^{-3}$ \citep{cr05b, jo10, ro07}.
Thus, they are important tracers during the prestellar and protostellar phase.

The N$_2$D$^+$/N$_2$H$^+$ abundance ratio can be an evolutionary indicator in early evolutionary stages of star formation \citep{cr05b, em09}.
The N$_2$D$^+$/N$_2$H$^+$ abundance ratio is affected by the exothermic reaction, H$_3^+ + $HD $\rightleftharpoons$ H$_2$D$^+ + $ H$_2$, since H$_2$D$^+$ transfers the deuteron to N$_2$ and produces N$_2$D$^+$ \citep{do04}.
Because CO is a destroyer of H$_2$D$^+$ and H$_3^+$ \citep{ca12}, the depletion of CO increases the abundance of H$_2$D$^+$ and H$_3^+$. The abundance increase of the latter molecule also speeds up the reaction to produce H$_2$D$^+$.
As a result, the CO depletion increases the abundance of H$_2$D$^+$ as well as N$_2$D$^+$ and leads to a high N$_2$D$^+$/N$_2$H$^+$ abundance ratio in a cold dense region.
\citet{cr05b} found that, in 31 low mass starless cores, the N$_2$D$^+$/N$_2$H$^+$ column density ratio (N(N$_2$D$^+$)/N(N$_2$H$^+$) ratio) is correlated with several evolutionary indicators such as core density, line width and line asymmetry, suggesting that the N(N$_2$D$^+$)/N(N$_2$H$^+$) ratio increases as a starless core evolves.
\citet{em09} showed that the N(N$_2$D$^+$)/N(N$_2$H$^+$) ratio decreases as the core evolves while getting hotter and less dense after the central protostar formed.
Thus, the N(N$_2$D$^+$)/N(N$_2$H$^+$) ratio increases in the prestellar core phase and declines in the protostellar core phase \citep{mi13,hu13}.

Infall motions are also used as an evolutionary indicator since they are likely to occur in the early stages \citep{em09}. Infall motions in protostars can be traced with observations including both optically thin and thick lines. 
For a protostar with an envelope dominated by infall, the foreground gas and background gas are red-shifted and blue shifted, respectively, with respect to the systemic velocity. If the foreground gas is colder than the central region, the red-shifted emission of an optically thick line can be absorbed, resulting in a spectral line asymmetry with a blue-shifted peak \citep{ma97}.

We study the chemical and physical properties of 16 Low Luminosity Objects (LLOs, $L_{\rm int}$ $\le$ 0.2 $L_\odot$) from \citet{du08}.
We describe our sample and observations in \S 2 and report the results in \S 3. The data analysis is presented in \S 4 and the properties of LLOs are discussed in \S 5.
We summarize the results in \S 6.

\section{OBSERVATIONS}
\subsection{Sample}
Using the c2d data \citep{ev03,ev09}, \citet{du08} identified 50 embedded protostars with $L_{\rm int}$ $\le$ 1 $L_\odot$ and 15 sources out of them are VeLLOs ($L_{\rm int}$ $\leq$ 0.1 $L_\odot$).
Hereafter we use the combination of the first letter of the three authors' last names plus the source number in their paper as our source name ({\it e.g.} DCE 185).
Our sample consists of 10 VeLLOs out of the 15 sources.
We excluded five sources: 1) DCE145 was identified as a background galaxy based on our CFHT Ks band (2.146 $\mu$m) outflow survey (Hsieh \& Lai in preparation), 2) DCE 161 and DCE 018 are unobservable using telescopes at the Arizona Radio Observatory (ARO), and 3) DCE 024 (CB-130) and DCE 032 (L1148) have spectra contaminated by their reference positions that were by mistake set too close to the sources.
In order to reduce the bias from the small sample, we added 6 Low Luminosity Objects (LLOs, 0.1 $L_{\odot}$ $<$ $L_{\rm int}$ $\leq$ 0.2 $L_{\odot}$) located in the two closest molecular clouds, Ophiuchus (125 pc) and Perseus (250 pc) from the \citet{du08} sample.
This sample includes 10 Class 0 sources and 6 Class I sources
(bolometric temperature $T_{\rm bol} >$ 70 K, Chen et al.\ 1995) as shown in Table 1.
These sources show evidence that they are indeed embedded, allowing us to study the chemical and physical properties of their parent cores.
\begin{deluxetable*}{ccccccc}
\tablewidth{0pt}
\tabletypesize{\tiny}
\tablecaption{LLO sample}
\tablehead{
\colhead{}    			&  \colhead{}   & \colhead{R.A.}  & \colhead{Dec}	& \colhead{$L_{\rm int}$}	& \colhead{Distance} & \colhead{$T_{\rm bol}$}	\\
\colhead{Source number}                       & \colhead{Other name} 		& \colhead{(J2000.0)}	& \colhead{(J2000.0)}	& \colhead{L$_{\odot}$}	& \colhead{(pc)} & \colhead{(K)}
}
\startdata
DCE 065		&			& 03:28:39.1		& 31:06:01.8		& 0.02	& 250 (50)	&	29 (3)\\
DCE 004		& L1521F	& 04:28:38.9		& 26:51:35.6		& 0.03	& 140 (10)	&	20 (3)\\
DCE 064		& 			& 03:28:32.6		& 31:11:05.3		& 0.03	& 250 (50)	&	65 (12)\\
DCE 031		& L673-7	& 19:21:34.8		& 11:21:23.4		& 0.04	& 300 (100)	&	24 (6)\\
DCE 001		& IRAM 04191 & 04:21:56.9	& 15:29:46.0		& 0.04	& 140 (10)	&	27 (3) \\
DCE181			&			& 16:28:48.5		& -24:28:38.6		& 0.05	& 125 (25)	& 	430 (19)\\
DCE 081		& 			& 03:30:32.7		& 30:26:26.5		& 0.06	& 250 (50)	&	33 (4)\\
DCE 025		& L328		& 18:16:59.5		& -18:02:30.5		& 0.07	& 270 (50)	&	62 (9)\\
DCE 038		& L1014		& 21:24:07.6		& 49:59:08.9		& 0.09	& 250 (50)	&	66 (21)\\
DCE 185		& IRAS 16253--2429 & 16:28:21.6	& -24:34:23.4		& 0.09	& 125 (25)	&	30 (2) \\
DCE 109		& 			& 03:44:21.4		& 31:59:32.6		& 0.11	& 250 (50)	&	345 (4)\\
DCE 092		& 			& 03:33:14.4		& 31:07:10.9		& 0.14	& 250 (50)	&	47 (5)\\
DCE 182		&			& 16:27:05.2		& -24:36:29.5		& 0.15	& 125 (25)	& 	105 (2)\\
DCE 107		& 			& 03:44:02.4		& 32:02:04.9		& 0.15	& 250 (50)	&	76 (4)\\
DCE 063		& 			& 03:27:38.3		& 30:13:58.8		& 0.20	& 250 (50)	&	199 (3)\\
DCE 090		& 			& 03:32:29.2		& 31:02:40.9		& 0.20	& 250 (50)	&	114 (17)
\enddata
\tablecomments{The source informations, Col. (3)-(7), are from \citet{du08}. The order of sources is based on their internal luminosities (Col. (5)) and is used in this paper.}
\end{deluxetable*}

\subsection{Kitt Peak 12m Telescope observations}
We observed N$_2$H$^+$ (1--0) toward the 16 LLOs  in November 2012 and February 2013 using the Kitt Peak 12 meter Telescope (KP12m) at the Arizona Radio Observatory (ARO).
We used the position-switching mode pointed toward the infrared sources \citep{du08} (Table 1).
The integration time (including on and off positions) for each source was 5--15 min.
The rms noise levels are in a range of 0.06 to 0.13 K at a spectral resolution of 0.078 km s$^{-1}$.
The half power beam width ($HPBW$) is $\sim$67.5\arcsec~and the pointing accuracy is 5\arcsec~rms.
The spectra were calibrated in main-beam temperature scale assuming a main-beam efficiency of 0.59 $\pm$ 0.07 based on the Saturn measurement (see ARO web page\footnote{http://aro.as.arizona.edu/12\_obs\_manual/appendix\_C.htm}).
We used the 3mm (ALMA Type Band 3) receiver and the backend Millimeter Autocorrelator with 6.2 kHz spectral resolution ($\sim$0.019 km s$^{-1}$), but the spectra were smoothed to 0.078 km s$^{-1}$ for the analysis.


\subsection{SMT observations}
The N$_2$H$^+$ (3--2), N$_2$D$^+$ (3--2), HCO$^+$ (3--2) and HCN$^+$ (3--2) observations were carried out between November 2012 and February 2014 using the Submillimeter Telescope (SMT) at the ARO, except for N$_2$D$^+$ (3--2) observation for DCE 065 which was carried out in December 2009.
The pointed positions were centered at the infrared sources (Table 1) and the observations were performed with the position-switching mode.
The integration time was 6--24 min for N$_2$H$^+$ (3--2), 30-42 min for N$_2$D$^+$ (3--2), 6 min for HCO$^{+}$ (3--2) and 5 min for HCN (3--2) for each source. 
The rms noise levels are 0.02--0.09 K for N$_2$H$^+$ (3--2), 0.01--0.04 K for N$_2$D$^+$ (3--2), 0.04--0.13 K for HCO$^{+}$ (3--2), and 0.05--0.08 K for HCN (3--2) at a spectral resolution of 250 kHz.
The $HPBWs$ are 27\arcsec~for N$_2$H$^+$ (3--2), 32.6\arcsec~for N$_2$D$^+$ (3--2) and $\sim$26\arcsec~for both HCO$^+$ (3--2) and HCN (3--2), and the pointing accuracy is 1\arcsec~rms.
We used the ALMA Type 1.3mm receiver and the filter-bank backend.
The main beam efficiency is 0.74 for all lines\footnote{http://aro.as.arizona.edu/smt\_docs/smt\_beam\_eff.htm}.
The spectral resolution is 250 kHz corresponding to 0.27 km s$^{-1}$ for N$_2$H$^+$ (3--2), 0.32 km s$^{-1}$ for N$_2$D$^+$ (3--2) and $\sim$0.28 km s$^{-1}$ for both HCO$^+$ (3--2) and HCN (3--2).

\section{RESULTS}

All the data were reduced using the CLASS program\footnote{http://www.iram.fr/IRAMFR/GILDAS/}. 
Figure 1 shows the spectra of N$_2$H$^+$ (1--0), N$_2$H$^+$ (3--2), N$_2$D$^+$ (3--2), HCO$^+$ (3--2) and HCN (3--2) for our targets.

\begin{figure*}
\includegraphics[scale=.45]{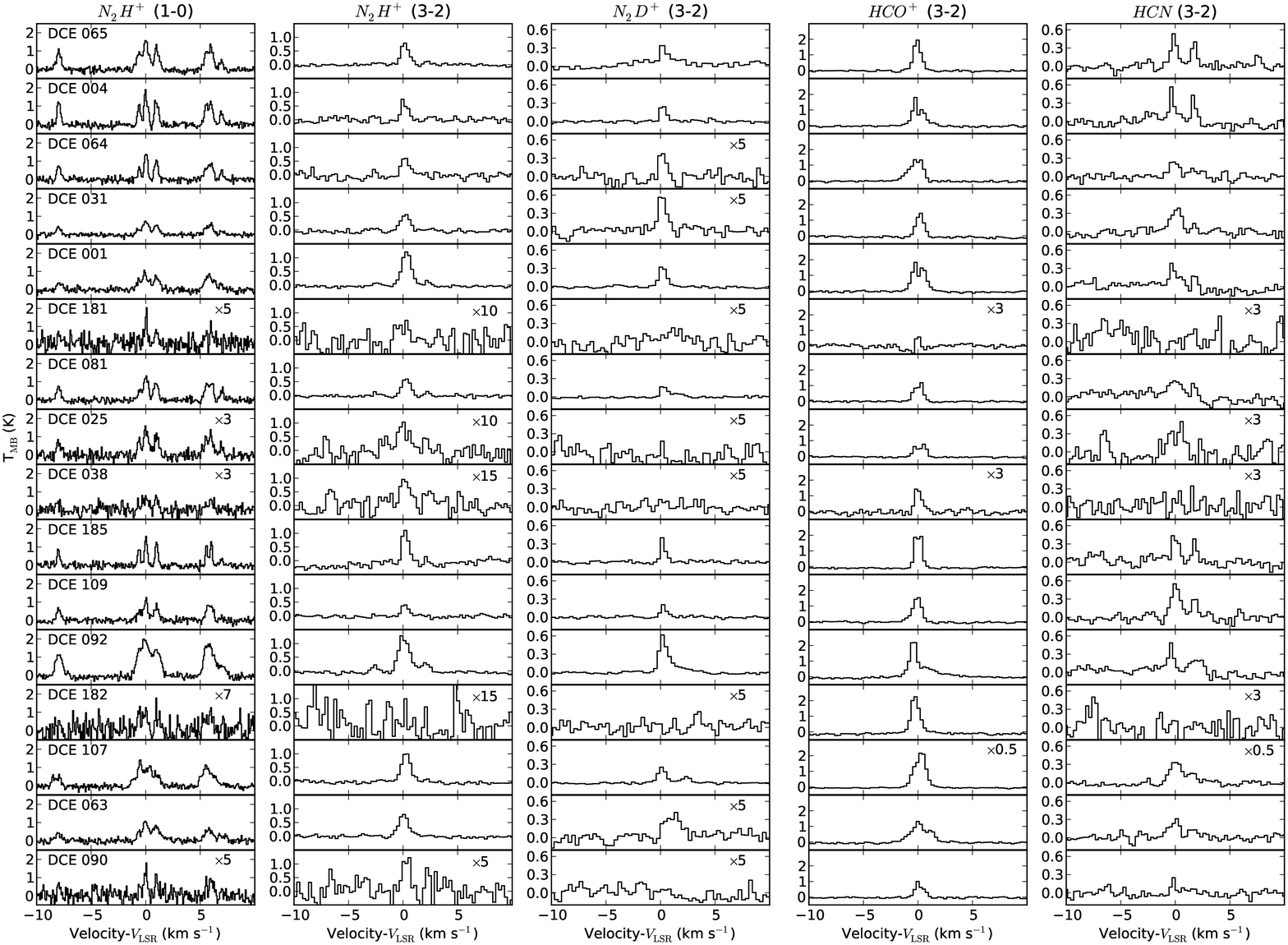}
\caption{The spectra of N$_2$H$^+$ (1--0), N$_2$H$^+$ (3--2), N$_2$D$^+$ (3--2), HCO$^+$ (3--2) and HCN (3--2) for all our LLO targets. The velocities are shifted by the $V_{\rm LSR}$ which is from the N$_2$H$^+$ (1--0) hyperfine fitting (Table 2).
The spectra of sources with weak or no detections are multiplied by a factor labeled in the upper right corner.
}
\end{figure*}

\subsection{N$_2$H$^+$ and N$_2$D$^+$}
\subsubsection{Spectra}

The N$_2$H$^+$ (1--0) lines were detected in all our targets although some sources have very weak detections.
We fitted all hyperfine components in N$_2$H$^+$ (1--0) by considering their relative intensities using our own code.
The code considers the relative intensities of all components similar to the fitting function ``METHOD HFS'' in CLASS.
The only difference is that we fit the excitation temperature ($T_{\rm ex}$) directly which gives us a direct estimate of its uncertainty whereas the uncertainty on $T_{\rm ex}$ needs to be obtained through error propagation from two parameters when using CLASS.
Since the ``METHOD HFS'' does not provide the covariance of the output parameters, our fitting process provides a more accurate error estimate for $T_{\rm ex}$.
Table 2 lists the best fitting results including systemic velocities ($V_{\rm LSR}$), line widths ($\Delta V$), optical depths ($\tau$) and excitation temperatures ($T_{\rm ex}$), together with integrated intensities (W).

\begin{deluxetable*}{cccccccccc}
\tablewidth{0pt}
\tabletypesize{\tiny}
\tablecaption{line properties} 
\tablehead{ 
\colhead{}    &  \multicolumn{5}{c}{N$_2$H$^+$ (1--0)}   & \multicolumn{2}{c}{N$_2$H$^+$ (3--2)} & \multicolumn{2}{c}{N$_2$D$^+$ (3--2)}
\\ 
\cmidrule(lr){2-6} \cmidrule(lr){7-8} \cmidrule(lr){9-10}
\colhead{}	& \colhead{W$_{N_2H^+_{J=1-0}}$} & \colhead{V$_\textmd{LSR}$}   & \colhead{$\Delta V$ (FWHM)}    & \colhead{$\tau_{\rm tot}$\tablenotemark{a}} & \colhead{$T_{\rm ex}$\tablenotemark{b}} &
\colhead{W$_{N_2H^+_{J=3-2}}$} & \colhead{$\tau_{\rm tot}$\tablenotemark{c}} &
\colhead{W$_{N_2D^+_{J=3-2}}$} & \colhead{$\tau_{\rm tot}$\tablenotemark{c}}
\\
\colhead{Source}	& \colhead{[K km s$^{-1}$]}	&\colhead{[km s$^{-1}$]}	&\colhead{[km s$^{-1}$]}	&\colhead{} &\colhead{[K]} 	&
\colhead{[K km s$^{-1}$]} &  &\colhead{[K km s$^{-1}$]}  &
}
\startdata
DCE 065 & 3.58 (0.07)   & 7.02 (0.01)   & 0.46 (0.02)   & 5.00 (1.13)   & 4.95 (0.30)   & 0.69 (0.04)   & 2.84 (1.10)     & 0.34 (0.04)   & $<$13.10\\
DCE 004 & 3.19 (0.08)   & 6.44 (0.01)   & 0.29 (0.01)   & 14.05 (2.22)  & 4.34 (0.09)   & 0.52 (0.06)   & $<$12.30& 0.19 (0.02)   & $<$6.60\\
DCE 064 & 2.31 (0.08)   & 7.19 (0.01)   & 0.35 (0.02)   & 2.94 (1.27)   & 5.52 (0.87)   & 0.56 (0.06)   & $<$18.70& 0.06 (0.02)   & $<$19.90\\
DCE 031 & 1.62 (0.08)   & 7.15 (0.01)   & 0.51 (0.03)   & 2.07 (1.30)   & 4.60 (0.91)   & 0.55 (0.04)   & $<$8.20& 0.11 (0.01)    & $<$14.00\\
DCE 001 & 2.28 (0.07)   & 6.63 (0.02)   & 0.62 (0.04)   & 3.05 (1.49)   & 4.39 (0.56)   & 1.29 (0.03)   & $<$2.40& 0.29 (0.01)    & $<$4.10\\
DCE 181 & 0.33 (0.06)   & 3.67 (0.01)   & 0.26 (0.03)   & 0.10 (0.07)   & 14.48 (18.46) & $<$0.02\tablenotemark{d}& -     & $<$0.02\tablenotemark{d}      & -\\
DCE 081 & 2.62 (0.07)   & 6.05 (0.01)   & 0.40 (0.02)   & 5.79 (1.44)   & 4.43 (0.24)   & 0.56 (0.03)   & 8.68 (2.28)     & 0.14 (0.01)   & $<$7.30\\
DCE 025 & 0.95 (0.07)   & 6.61 (0.01)   & 0.40 (0.03)   & 4.43 (2.12)   & 3.51 (0.24)   & 0.11 (0.02)   & $<$19.90& $<$0.02\tablenotemark{d}      & -\\
DCE 038 & 0.40 (0.05)   & 4.24 (0.03)   & 0.44 (0.07)   & 1.95 (4.19)   & 3.40 (1.16)   & 0.07 (0.02)   & $<$19.90& $<$0.03\tablenotemark{d}      & -\\
DCE 185 & 2.10 (0.05)   & 4.04 (0.004)   & 0.23 (0.01)   & 5.76 (1.51)   & 4.93 (0.34)   & 0.73 (0.03)   & $<$19.50& 0.21 (0.01)   & $<$4.90\\
DCE 109 & 2.07 (0.07)   & 9.12 (0.01)   & 0.45 (0.02)   & 0.87 (1.31)   & 8.49 (7.59)   & 0.32 (0.04)   & $<$12.10& 0.14 (0.02)   & $<$7.50\\
DCE 092 & 6.46 (0.08)   & 6.27 (0.01)   & 0.66 (0.02)   & 6.89 (0.88)   & 5.12 (0.14)   & 1.48 (0.05)   & 5.06 (1.33)     & 0.59 (0.02)   & $<$0.90\\
DCE 182 & 0.32 (0.05)   & 4.84 (0.03)   & 0.37 (0.07)   & 5.55 (6.20)   & 2.97 (0.16)   & $<$0.03\tablenotemark{d}& -     & $<$0.02\tablenotemark{d}      & -\\
DCE 107a        & 1.34 (0.85)   & 8.95 (0.02)   & 0.62 (0.06)   & 1.57 (0.46)   & 4.61 (0.76)   & 1.04 (0.04)   & $<$3.90 & 0.23 (0.02)   & $<$4.50\\
DCE 107b        & 0.87 (1.66)   & 8.40 (0.01)   & 0.33 (0.03)   & 1.08 (0.87)   & 6.30 (4.45)   & 1.04 (0.04)   & $<$3.90 & 0.23 (0.02)   & $<$4.50\\
DCE 063 & 2.59 (0.08)   & 4.71 (0.02)   & 0.65 (0.04)   & 1.35 (1.31)   & 5.92 (2.53)   & 0.84 (0.04)   & $<$4.50& 0.13 (0.02)    & $<$19.90\\
DCE 090 & 0.37 (0.04)   & 6.62 (0.01)   & 0.29 (0.03)   & 0.10 (1.79)   & 15.15 (325.44)        & 0.21 (0.03)   & $<$19.90        & $<$0.02\tablenotemark{d}      & -
\enddata
\tablecomments{The values in parenthesis represent 1$\sigma$ uncertainties.}
\tablenotetext{a}{$\tau$ represents the total opacity which is the sum of the optical depth of all hyperfine components. The fitting limitation of $\tau$ is set to be $\tau$ $>$ 0.1.}
\tablenotetext{b}{The excitation temperatures are computed by assuming a beam filling factor of 1.}
\tablenotetext{c}{The upper limits of the total opacity $\tau$ are estimated based on the peak intensities and the rms noise levels since most of the satellite lines are undetected. }
\tablenotetext{d}{The integrated intensity upper limits is estimated using a peak temperature detection limit of 3$\sigma$, and assuming the line width is the same as N$_2$H$^+$ (1--0).}
\end{deluxetable*}

The N$_2$H$^+$ (3--2) lines were detected toward all the targets except DCE 181 and DCE 182 which have weak detections in N$_2$H$^+$ (1--0).
The N$_2$D$^+$ (3--2) lines were detected toward 11 targets and were undetected in DCE 181, 025, 038, 182 and 090.
The satellite lines are not clearly detected with the rms noise $\sigma$ of 0.02--0.09 K for N$_2$H$^+$ (3--2) and 0.01-0.04 K for N$_2$D$^+$ (3--2). 
This prevents us from deriving the excitation temperatures and optical depths (for most of our targets) with HFS fitting.
However, we can estimate the upper limits of $\tau$ for each source.
If we use a threshold 3$\sigma$ as the upper limit of the undetected satellite lines, we obtain a corresponding upper limit of $\tau$.
(The optical depths of N$_2$H$^+$ (3--2) of DCE 065, 81 and 092 are derived by fitting the hyperfine structure with relative intensity.)
We thus list the integrated intensities and upper limits of optical depths for N$_2$D$^+$ and N$_2$H$^+$ (3--2) in Table 2.

\subsubsection{Two components of DCE 107}
The N$_2$H$^+$ (1--0) spectrum of DCE 107 shows a clear double peak structure.
The SCUBA 850 $\mu$m continuum emission image from the COMPLETE project \citep{ri06} has resolved two components with a separation of $\sim$ 35\arcsec~within our observing beam.
Therefore, we suspect that the two velocity components seen in the N$_2$H$^+$ (1--0) spectrum correspond to the two continuum sources in the 850 $\mu$m image (hereafter DCE 107a and DCE 107b).
We fitted the spectrum using a superposition of two sets of hyperfine components; each set has its own systemic velocity, line width, optical depth and excitation temperature (Table 2). 
The fitting result seems to be reasonable by comparing with the N$_2$H$^+$ (1--0) observations from \citet{ki07}.
Using the IRAM 30m telescope ($\theta_{FWHM} \sim$ 27\arcsec), \citet{ki07} observed DCE 107 (core 21 in their paper) at a position $\sim$12\arcsec~away from our position (the infrared source) and $\sim$8\arcsec~away from DCE107a.
They detected only one velocity component, supporting our interpretation above, and they obtained a line width of 0.65 $\pm$ 0.009 km s$^{-1}$ and a systemic velocity of 9.02 $\pm$ 0.008 km s$^{-1}$ which are very close to the values we derive for DCE 107a ($\Delta V$ $=$ 0.61 $\pm$ 0.06 km s$^{-1}$ and $V_{\rm LSR}$ $=$ 8.95 $\pm$ 0.02 km s$^{-1}$).


\subsection{HCO$^+$ and HCN observations}
\subsubsection{spectra}
The HCO$^+$ (3--2) line was detected toward all 16 LLO targets and the HCN (3--2) line toward 13 out of the 16 targets (undetected in DCE 181, 038 and 182). We use HCO$^{+}$ (3--2) and HCN (3--2) as our optically thick tracers to search for infall signatures. The optical depth of HCN (3--2) was estimated by fitting the hyperfine components. 
The optical depths of our targets are all larger than 5 (except for DCE 025 and DCE 090 which have very uncertain values), 
indicating that the HCN (3--2) lines are optically thick toward our targets (Table 3).
This feature can be easily seen in the spectra since the satellite lines (F$=$ 3 $\rightarrow$ 3) are significant compared with the main components.
Note that the satellites F $=$ 2 $\rightarrow$ 2 are much weaker or undetected although F $=$ 2 $\rightarrow$ 2 and F $=$ 3 $\rightarrow$ 3 have equal line strengths (anomalies, Daniel et al. 2006).

\begin{deluxetable*}{cccccccc}
\tablewidth{0pt}
\tabletypesize{\tiny}
\tablecaption{Line properties of HCO$^{+}$ (3--2) and HCN (3--2)} 
\tablehead{ 
\colhead{}    & \colhead{HCO$^{+}$ (3--2)} &  \multicolumn{5}{c}{HCN (3--2)}  
\\ 
\cmidrule{3-7}
\colhead{}	& \colhead{W$_{HCO^+_{J=3-2}}$} & \colhead{V$_\textmd{LSR}$}   & \colhead{$\Delta V$}    & \colhead{$\tau_{\rm tot}$\tablenotemark{b}} & \colhead{$T_{\rm ex}$\tablenotemark{c}} &
\colhead{W$_{HCN_{J=3-2}}$} 
\\
\colhead{Source}	& \colhead{[K km s$^{-1}$]}	&\colhead{[km s$^{-1}$]}	&\colhead{[km s$^{-1}$]}	&\colhead{} &\colhead{[K]} 	&\colhead{[K km s$^{-1}$]} 
}
\startdata
DCE 065 & 5.96 (0.03)   & 6.92 (0.01)   & 0.27 (0.03)   & 28.04 (9.85)  & 3.98 (0.10)   & 2.37 (0.04)  \\
DCE 004 & 6.52 (0.07)   & 6.30 (0.02)   & 0.37 (0.04)   & 43.28 (18.33) & 3.75 (0.09)   & 1.99 (0.07)  \\
DCE 064 & 7.25 (0.05)   & 7.05 (0.05)   & 0.47 (0.09)   & 13.19 (7.46)  & 3.50 (0.09)   & 1.43 (0.06)  \\
DCE 031 & 4.37 (0.05)   & 7.17 (0.05)   & 0.72 (0.13)   & 5.16 (3.04)   & 3.82 (0.11)   & 1.81 (0.07)  \\
DCE 001 & 8.23 (0.07)   & 6.66 (0.05)   & 0.56 (0.07)   & 19.64 (8.84)  & 3.54 (0.08)   & 1.50 (0.07)  \\
DCE 181 & 0.19 (0.02)   & -     & -     & -     & -     & -      \\
DCE 081 & 4.00 (0.04)   & 6.01 (0.05)   & 0.57 (0.07)   & 29.90 (13.24) & 3.54 (0.09)   & 1.83 (0.05)  \\
DCE 025 & 2.93 (0.04)   & 6.71 (0.21)   & 1.30 (1.05)   & 1.21 (7.07)   & 3.34 (1.68)   & 0.60 (0.06)  \\
DCE 038 & 1.36 (0.03)   & -     & -     & -     & -     & -      \\
DCE 185 & 6.10 (0.07)   & 4.05 (0.03)   & 0.32 (0.05)   & 36.18 (15.39) & 3.87 (0.10)   & 2.01 (0.09)  \\
DCE 109 & 5.63 (0.03)   & 9.13 (0.03)   & 0.54 (0.05)   & 17.57 (5.14)  & 3.97 (0.08)   & 2.73 (0.04)  \\
DCE 092 & 10.06 (0.10)  & 5.82 (0.04)   & 0.44 (0.07)   & 12.74 (6.40)  & 3.76 (0.10)   & 2.36 (0.11)  \\
DCE 182 & 7.63 (0.06)   & -     & -     & -     & -     & -      \\
DCE 107 & 19.73 (0.07)  & 8.91 (0.03)   & 0.73 (0.06)   & 8.20 (2.29)   & 4.37 (0.07)   & 4.38 (0.10)  \\
DCE 063 & 8.18 (0.05)   & 4.45 (0.05)   & 0.67 (0.08)   & 14.09 (5.77)  & 3.51 (0.07)   & 1.73 (0.05)  \\
DCE 090 & 3.03 (0.04)   & 6.42 (0.19)   & 0.29 (0.55)   & 0.10 (3.50)   & 9.32 (174.78) & 0.67 (0.04) 
\enddata
\tablecomments{The values in parenthesis represent 1$\sigma$ uncertainties.}
\tablenotetext{a}{$\tau$ represents the total opacity which is the sum of the optical depth of all hyperfine components. The fitting limitation of $\tau$ is set to be $\tau$ $>$ 0.1.}
\tablenotetext{b}{The excitation temperatures are computed by assuming a beam filling factor of 1.}
\end{deluxetable*}

\section{ANALYSIS}

\subsection{N$_2$H$^+$ Non-LTE RADEX analysis}
We used the non-LTE radiative transfer code RADEX \citep{va07} to model the N$_2$H$^+$ (1--0) and (3--2) spectra for our targets.
The densities of low mass protostellar cores can be lower than the critical densities of N$_2$H$^+$ (1--0) and (3--2).
This means that the local thermal equilibrium (LTE) assumption is not appropriate for deriving the true physical properties.
RADEX provides fast non-LTE analysis for molecular line spectra in a uniform medium, which involves collisional and radiative processes.
The N$_2$H$^+$ line data (energy levels, statistical weights, Einstein A-coefficients and collisional rate coefficients) are from \citet{da05} and \citet{sc05}, and were taken from the Leiden Atomic and Molecular Database (LAMDA).

We used RADEX to derive the cores' kinetic temperatures ($T_{\rm kin}$), N$_2$H$^+$ column densities (N(N$_2$H$^+$)), H$_2$ densities ($n_{\rm H_2}$) and source sizes.
The former three parameters are related to the cores' evolutionary state and/or may decide the mass of stars formed in the future. 
The core sizes, together with the beam sizes, yield the beam filling factors that determine the scale of the spectra.
Since the beam sizes of our N$_2$H$^+$ (1--0) and (3--2) observations are different, the beam filling factors affect the intensity ratios of N$_2$H$^+$ (1--0)/(3--2) as well as our derived physical parameters. 

We constructed a model grid for each source to derive the physical parameters.
The grid is composed with three variances, kinetic temperature, N$_2$H$^+$ column density and H$_2$ density, while the line width is fixed to that from the hyperfine fitting. 
The step sizes of the grid are 0.5 K for $T_{\rm kin}$, 0.025 and 0.05 in decimal log scale for N(N$_2$H$^+$) and $n_{\rm H_2}$, respectively.
Each cell in the grid contains the synthetic spectra of N$_2$H$^+$ (1--0) and (3--2).
The synthetic spectra are built-up based on RADEX results which include the excitation temperatures ($T_{\rm ex}$) and optical depths ($\tau$) for all hyperfine components. The spectra are constructed using the equation
\begin{equation}
T_{\rm MB} (v) = \Psi (\frac{\sum J(T_{\rm ex}^{i}) \tau_{i}(v)}{\sum \tau_{i}(v)} -J(T_{\rm bg}))(1-\exp(-\sum \tau_{i}(v))),
\end{equation}
where $\Psi$ is the beam filling factor, $T_{\rm bg}$ is the cosmic background temperature (2.73 K), and $J(T) = [h\nu/k]/[\exp(h\nu/kT)-1]$. 
To approach a reasonable intensity at the hyperfine component overlapping region, we used a weighting emissivity term, $\frac{\sum J(T_{\rm ex}^{i}) \tau_{i}(v)}{\sum \tau_{i}(v)}$. 
This conception is similar to the averaged excitation temperature in \citet{da06} for discussing the line anomalies ({\it{i.e.}} $T_{\rm ex}$ are different between the hyperfine components.).
The weighting is based on the optical depths contributed from all components at a specific velocity.
$\sum \tau_{i}(v)$ is the superposition of optical depths contributed from all components at such velocity.
Note that RADEX does not consider the line overlap, which may make our approach uncertain at such regions \citep{da06,va07}.
The beam filling factor ({\it i.e.} source size) is the free parameter determining the scale of the synthetic spectra.
As a result, we obtained a $\chi^2$ distribution (see Appendix and Figure A1) and the corresponding best fitting core size distribution in the grid for each source.
Figure A2 shows an example (DCE 081) of the best fitting results using our RADEX synthetic spectra compared with using HFS fitting.
The RADEX synthetic spectra seem to fit the hyperfine anomalies better than HFS fitting for the 121-011 and 122-011 components. For the 112-012 component, both fitting processes can not fit the anomalies well. For most of our sources, no significant difference is found between the two fitting processes.
The best fitting results are listed in Table 4 for the 13 targets with both N$_2$H$^+$ (1--0) and (3--2) detections and only one velocity component.
The errors are defined with a 68.3\% confidence level area \citep{pr92} for four free parameters ($T_{\rm kin}$, N(N$_2$H$^+$), $n_{\rm H_2}$ and core sizes).

\begin{deluxetable}{ccccc}
\tablewidth{0pt}
\tabletypesize{\tiny}
\tablecaption{RADEX fitting results} 
\tablehead{ 
\colhead{}	& \colhead{$T_{\rm kin}$} & \colhead{$n_{\rm H_2}$}   & \colhead{N(N$_2$H$^+$)}    & \colhead{FWHM}
\\
\colhead{Source}	& \colhead{[K]}	&\colhead{log$_{10}$ [cm$^{-3}$]}	&\colhead{log$_{10}$ [cm$^{-2}$]}	&\colhead{[arcsec]}
}
\startdata
DCE065  & 7.5$_{-1.5}^{+2.5}$   & 6.1$_{-0.4}^{+0.6}$   & 13.10$_{-0.12}^{+0.07}$       & 58$_{-12}^{+27}$\\
DCE004  & 8.0$_{-\infty}^{+3.5}$   & 5.8$_{-0.4}^{+\infty}$        & 13.35$_{-0.30}^{+0.15}$       & 45$_{-9}^{+66}$\\
DCE064  & 13.5$_{-5.5}^{+8.0}$  & 5.4$_{-0.3}^{+0.6}$   & 13.02$_{-0.24}^{+0.18}$       & 42$_{-8}^{+25}$\\
DCE031  & 25.5$_{-13.0}^{+\infty}$      & 5.0$_{-\infty}^{+0.5}$   & 13.23$_{-0.21}^{+0.14}$       & 27$_{-4}^{+7}$\\
DCE001  & 6.5$_{-0.0}^{+1.0}$    & 7.3$_{-0.7}^{+0.2}$   & 13.17$_{-0.07}^{+0.06}$       & 42$_{-3}^{+3}$\\
DCE081  & 16.5$_{-5.5}^{+6.0}$  & 5.2$_{-0.2}^{+0.3}$   & 13.32$_{-0.09}^{+0.13}$       & 34$_{-3}^{+4}$\\
DCE025  & 6.0$_{-\infty}^{+9.0}$   & 5.6$_{-1.0}^{+\infty}$        & 12.82$_{-0.62}^{+0.31}$       & 40$_{-11}^{+\infty}$\\
DCE038  & 9.0$_{-\infty}^{+\infty}$        & 5.0$_{\infty}^{+\infty}$      & 12.95$_{-1.35}^{+0.55}$       & 24$_{-9}^{+\infty}$\\
DCE185  & 6.5$_{\infty}^{+3.0}$    & 7.3$_{-1.3}^{+\infty}$        & 12.78$_{-0.13}^{+0.22}$       & 77$_{-34}^{+18}$\\
DCE109  & 7.5$_{-\infty}^{+15.5}$  & 6.0$_{-1.0}^{+\infty}$        & 12.63$_{-0.31}^{+0.35}$       & 85$_{-40}^{+\infty}$\\
DCE092  & 7.5$_{-0.5}^{+0.5}$   & 6.3$_{-0.1}^{+0.2}$   & 13.37$_{-0.02}^{+0.06}$       & 65$_{-7}^{+7}$\\
DCE063  & 33.0$_{-25.5}^{+\infty}$      & 5.0$_{-\infty}^{+1.4}$   & 13.23$_{-0.33}^{+0.09}$       & 35$_{-4}^{+19}$\\
DCE090  & 30.5$_{-21.0}^{+\infty}$      & 5.2$_{-\infty}^{+0.8}$   & 13.28$_{-0.40}^{+0.29}$       & 11$_{-3}^{+6}$
\enddata
\tablecomments{The parameters of the best-fit model are listed in the table. The errors presented here are with 68.3\% confidence level. The $\infty$ error indicates that the parameter is not constrained at the upper or lower end.}
\end{deluxetable}

Out of the 13 targets, $T_{\rm kin}$ and $n_{\rm H_2}$ are constrained in 5 sources but not in 8 sources (which have a confidence interval exceeding the boundary of our model grid, see Figure A1). For the 5 sources with constraints, the average error is $\sim$3.1 K for $T_{\rm kin}$ and a multiplicative factor of $\sim$0.36 for $n_{\rm H_2}$ (steps by log scale).
DCE 064 has the most uncertain $T_{\rm kin}$ which is 13.5$_{-5.5}^{+8.0}$ K and DCE 065 has the most uncertain $n_{\rm H_2}$ which is 10$^{6.1_{-0.4}^{+0.6}}$ cm$^{-3}$.
For the 8 unconstrained sources, we list the parameters from the best-fit model and upper or lower limits of these parameters (Table 4).

We here compare the H$_2$ densities of our fitting results with the literature for several well-studied objects.
Our derived densities are quite consistent with that of DCE 004 (L1521F) and 038 (L1014) in \citet{cr04,cr05a}.
Considering the 1.2 mm continuum maps and a density profile, $n_{\rm H_2}$ $=$ $n_0$/(1+$r/r_0$)$^\alpha$, they found $n_0$ $=$ 10$^{6}$ cm$^{-3}$ and $\alpha \sim 2$ for DCE 004, and $n_0$ $=$ 2.5 $\times$ 10$^{5}$ cm$^{-3}$ and $\alpha \sim 2.7$ for DCE 038 (both with a ``flat'' region, r$_0$ $\sim$ 20\arcsec). Our derived densities (6.3 $\times$ 10$^5$ cm$^{-3}$ within r $<$ 23\arcsec~for DCE 004, and 10$^5$ cm$^{-3}$ within r $<$ 12\arcsec~for DCE 038) are very close to the results from the literature (but DCE 038 is not well constrained in our sample).
However, the particularly high densities of DCE 001 (IRAM 04191) and DCE 185 (IRAS 16253) are very likely overestimated.
For DCE 001, \citet{be02} derived a density of $\sim 2 \times 10^5$ 
cm$^{-3}$ at a radius of 20\arcsec, based on the dust continuum emission and the 
radiative transfer modeling of CS transitions.
Our estimated density, 2 $\times$ 10$^7$ cm$^{-3}$ within a radius of $\sim$21\arcsec~is obviously too high.
DCE 001 has the highest N$_2$H$^+$ (3--2)/(1--0) intensity ratio 
among our targets (Table 2). In addition, our observed 
N$_2$H$^+$ (1--0) peak intensity is lower than that from the IRAM 30m 
telescope \citep{be02} by a factor of $\sim$3. The optical depth derived from 
the HFS fitting is also lower than that in \citet{be02} by a factor of 
$\sim$2.
We therefore speculate that our observation with a large beam ($HPBW =$ 67.5\arcsec) is seriously contaminated by diffuse regions whereas the IRAM 30m observation ($HPBW$ $\sim$27\arcsec) is not.
Our N$_2$H$^+$ (3--2) observations have the same angular resolution as
the IRAM 30m observations of N$_2$H$^+$ (1--0), so they likely trace the same
region. The contamination of our N$_2$H$^+$ (1--0) data by more diffuse 
material likely leads to an overestimate of the N$_2$H$^+$ (3--2)/(1--0) 
intensity ratio that is used in our analysis to derive constraints on the 
density. Taking into account beam filling factor effects as we did in our
analysis cannot solve this problem.
The case of DCE 185 is very similar. 
The high N$_2$H$^+$ (3--2)/(1--0) intensity ratio is likely due to
the contamination of our 1-0 spectrum by more diffuse regions;
the high density (2 $\times$ 10$^7$ cm$^{-3}$) within a radius of 39\arcsec~are unlikely to be true since the envelope mass of DCE 185 is small (0.15--1.0 $M_\odot$ Barsony et al.\ 2010, Stanke et al.\ 2006, Enoch et al.\ 2008 and Tobin et al.\ 2012).
Some sources with high N$_2$H$^+$ (3--2)/(1--0) intensity ratios may also suffer from this problem such as DCE 031 and 063 (intensity ratios $\sim$ 3).
As a result, these two sources have the highest kinetic temperatures among our targets for explaining their high intensity ratios in our RADEX fittings.
The kinetic temperatures could thus be overestimated.
It is likely that sources with relatively low N$_2$H$^+$ (3--2)/(1--0) intensity ratios (and/or high opacities) are less affected by the contamination of diffuse regions.
N$_2$H$^+$ (3--2) and (1--0) observations with a similar beam size are required for deriving the densities of DCE 001, 185, 031, and 063
in a more reliable way.

\subsection{N$_2$D$^+$/N$_2$H$^+$ column density ratio}
The N$_2$D$^+$ column density is difficult to derive with only the J $=$ 3 $\rightarrow$ 2 transition because we cannot derive the excitation temperature ($T_{\rm ex}$) and optical depth ($\tau$) which are crucial for determining the column density (see \S 3.1.1).
In addition, the excitation temperatures could be very different between different transitions under non-LTE conditions (see Fig.\ 17 in Daniel et al.\ 2007).
Using N$_2$H$^+$ as an example, the excitation temperatures of N$_2$H$^+$ (3--2) are lower than that of N$_2$H$^+$ (1--0) by 0.3 -- 4 K in our RADEX best-fit models (see \S 4.1).

With only one transition, we have to perform an LTE analysis to derive the column density of N$_2$D$^+$. However,
a reasonable N$_2$D$^+$/N$_2$H$^+$ column density ratio (N(N$_2$D$^+$)/N(N$_2$H$^+$)) can still be obtained considering that 
\citet{da07} found very similar excitation temperatures for N$_2$D$^+$ and N$_2$H$^+$ in a given transition (J $=$ 1 $\rightarrow$ 0 or J $=$ 2 $\rightarrow$ 1 or J $=$ 3 $\rightarrow$ 2, see their Fig.\ 17).
For the J $=$ 3 $\rightarrow$ 2 transition, the largest difference in excitation temperature is $\lesssim$ 0.5 K at a H$_2$ density of $\sim$ 10$^{5}$ cm$^{-3}$.
We therefore assume that the N(N$_2$D$^+$)/N(N$_2$H$^+$) ratio derived using the same transition (J $=$ 3 $\rightarrow$ 2) in both molecules will be very close to the actual ratio, even when deriving each column density with LTE equations.
We use the excitation temperatures of N$_2$H$^+$ (3--2) obtained from the RADEX best-fit model for both N$_2$H$^+$ and N$_2$D$^+$ (3--2).
(Since each hyperfine component of N$_2$H$^+$ (3--2) has a different $T_{\rm ex}$, we use the averaged excitation temperature as defined in \citet{da06}, $T_{\rm ave} = \frac{h\nu}{k}/\ln[1+\frac{\frac{h\nu}{k}\sum\tau_{i}}{\sum\tau_{i}J(T^{i}_{\rm ex})}]$.)
The opacity of N$_2$D$^+$ (3--2) is then calculated by fitting the hyperfine components with $T_{\rm ex}$ fixed to the value derived for N$_2$H$^+$ (3--2). Note the N$_2$D$^+$ (3--2) spectra were scaled with the beam filing factors derived from source sizes.
The column densities are derived using Eq.\ (76) in \citet{ma13} and equation A4 in \citet{ca02},
\begin{equation}
N^{\rm thick}_{\rm tot} = N^{\rm thin}_{\rm tot}\frac{\tau}{1-exp(-\tau)}
\end{equation}
and
\begin{align}
&N^{\rm thin}_{\rm tot} = \frac{8\pi W}{\lambda^3A}\frac{g_{l}}{g_{u}} \frac{1}{[J(T_{\rm ex})-J(T_{\rm bg})]}   \nonumber \\
&\frac{1}{1-\exp(-h\nu/kT_{\rm ex})}   \frac{Q_{\rm rot}(T_{\rm ex})}{g_l \exp(-E_{l}/k T_{\rm ex})}	\nonumber \\
\end{align}
where $\tau$ is the total opacity of the multiplet, $W$ is the total integrated intensity of the multiplet (scaled by the beam filling factor $\theta_{\rm size}^2/(\theta_{\rm size}^2+\theta_{\rm beam}^2)$, see \S 4.1), $A$ is the Einstein coefficient, $g_l$ and $g_u$ are the statistical weights of the lower and upper levels, $Q_{\rm rot}$ is the rotational partition function under LTE conditions, and $E_{l}$ is the energy of the lower level.
The column densities are listed in Col. (3) and Col. (4) of Table 5 and the N(N$_2$D$^+$)/N(N$_2$H$^+$) ratios are listed in Col. (5).
The uncertainties reported in Col. 5 take into account the uncertainties on the measured integrated intensities but neglect the uncertainty on $T_{\rm ex}$ because it does not significantly affect the N(N$_2$D$^+$)/N(N$_2$H$^+$) ratio.
For a typical excitation temperature of $\sim$ 5.4 K in our sample, the N(N$_2$D$^+$)/N(N$_2$H$^+$) ratio will only increase by $\sim$ 7\% and decrease by $\sim$ 9\% if the excitation temperature increases by 1 K and decreases by 1 K, respectively. The effect is even smaller at a higher excitation temperature; the N(N$_2$D$^+$)/N(N$_2$H$^+$) ratio will increase by $\sim$20\% if the excitation temperature increases by 5 K. 
Another noteworthy uncertainty could come from the unresolved hyperfine structure of J $=$ 3 $\rightarrow$ 2.
We note also that the opacity correction applied in Eq. (2) slightly overestimates the column density because the total opacity of the multiplet, $\tau$,
is higher than the opacities of the individual components.

\begin{deluxetable}{ccccp{2.1cm}}
\tabletypesize{\tiny}
\tablecaption{N$_2$D$^+$/N$_2$H$^+$ column density ratios} 
\tablehead{ 
\colhead{}	& \colhead{N(N$_2$H$^+$)} & \colhead{N(N$_2$H$^+$)}  & \colhead{N(N$_2$D$^+$)}   & \colhead{N(N$_2$D$^+$)/N(N$_2$H$^+$)}  
\\
\colhead{Source}	& \colhead{10$^{12}$ cm$^{-2}$} & \colhead{10$^{12}$ cm$^{-2}$}	&\colhead{10$^{11}$ cm$^{-2}$}	&\colhead{}
}
\startdata
DCE 065	& 16.73 (0.96)	& 27.54 (1.39)	& 23.91 (2.81)	& 0.087 (0.011)	\\
DCE 004	& 33.94 (2.24)	& 88.26 (10.88)	& 23.96 (2.92)	& 0.027 (0.005)	\\
DCE 064	& 14.15 (1.69)	& 27.95 (3.20)	& 5.03 (1.46)	& 0.018 (0.006)	\\
DCE 031	& 21.38 (3.38)	& 28.53 (2.18)	& 11.43 (1.63)	& 0.040 (0.006)	\\
DCE 001	& 8.53 (1.41)	& 24.96 (0.54)	& 13.97 (0.40)	& 0.056 (0.002)	\\
DCE 081	& 28.96 (2.43)	& 54.71 (0.27)	& 15.50 (0.82)	& 0.028 (0.002)	\\
DCE 025	& 6.02 (1.85)	& 14.04 (2.76)	& -	& -	\\
DCE 185	& 6.72 (0.48)	& 11.83 (0.48)	& 7.55 (0.46)	& 0.064 (0.005)	\\
DCE 109	& 5.04 (0.60)	& 6.23 (0.69)	& 10.58 (1.21)	& 0.170 (0.027)	\\
DCE 092	& 25.93 (1.06)	& 50.59 (1.68)	& 30.68 (0.86)	& 0.061 (0.003)	\\
DCE 063	& 16.75 (2.30)	& 24.12 (1.10)	& 9.41 (1.34)	& 0.039 (0.006)	
\enddata
\tablecomments{Col.\ (1): Source name. Col.\ (2)-(3): N$_2$H$^+$ column density derived from N$_2$H$^+$ (1--0) observations and N$_2$H$^+$ (3--2) observations, respectively (see \S 4.2 for details). Col.\ (4): N$_2$D$^+$ column density derived from N$_2$D$^+$ (3--2) (see text for details). Col.\ (5): Ratio of Col. (4) and (3).}
\tablecomments{The values in parenthesis represent 1$\sigma$ uncertainties.}
\tablecomments{DCE 038 and DCE 090 column densities are not derived because their N$_2$H$^+$ 101-012 component is not detected (see text for details).}
\end{deluxetable}

We also derive the N$_2$H$^+$ column densities from the J $=$ 1 $\rightarrow$ 0 transition with Eqs. (2) and (3), using only
the 101-012 component that is not affected by other hyperfine components (see Col. (2) of Table 5).
The N$_2$H$^+$ column densities  derived from N$_2$H$^+$ (1--0) are slightly larger than that from RADEX (the best-fit models) by factors of 1.0 to 1.5 except for DCE 001 (0.6) and DCE 025 (0.9).
The average of the factors is 1.2 with a standard deviation of 0.25.
This result makes sense because LTE analysis will overestimate the N$_2$H$^+$ populations in J $> 1$ levels when the collision rate is insufficient to pump N$_2$H$^+$ with low volume densities (non-LTE).
The column densities derived from N$_2$H$^+$ (3--2) (Col. (3) of Table 5) are also larger than that from the RADEX best-fit model (Col. (4) of Table 4) by factors of 1.4 -- 3.9, with an average value of 2.1 and a standard deviation of 0.6. The column densities of N$_2$D$^+$ listed in Col. (4) of Table 5 are thus likely overestimated by similar factors.
However, as explained above, the N$_2$D$^+$/N$_2$H$^+$ column density ratios in Col. (5) should be reliable.

\subsection{Asymmetry parameters}

To probe the infall motion of our target cores, we used the asymmetry parameter defined by \citet{ma97},
\begin{equation}
\delta v = \frac{v_\textmd{thick}-v_\textmd{thin}}{\Delta v_\textmd{thin}},
\end{equation}
where $v_\textmd{thick}$ and $v_\textmd{thin}$ are the velocities of the optically thick line and optically thin line, respectively, and $\Delta v_\textmd{thin}$ is the line width of the optically thin line.
We used N$_2$H$^+$ (1--0) as our optically thin tracer, and $v_\textmd{thin}$ and $\Delta v_\textmd{thin}$ are derived from HFS fitting (Table 2).
The HFS fitting provides the line width that corresponds to the width of the opacity profile. It should
represent the velocity dispersion well and can thus be used as a proxy for $\Delta v_\textmd{thin}$.
Eleven out of our sixteen sources have optical depths lower than 1.3 for the strongest hyperfine component (123--012, $\tau_{\rm tot}~\times~7/27$), suggesting that the systemic velocity derived from the HFS fitting should not be significantly affected by optical depth effects and can be used as a proxy for $v_\textmd{thin}$. For the remaining sources, three (DCE 081, 185, and 092) have optical depths lower than 1.3 for the isolated hyperfine component (101--012, $\tau_{\rm tot}~\times~3/27$). The systemic velocity derived from the HFS fitting matches well the centroid velocity of the isolated component for these three sources and can thus also be used as a proxy for $v_\textmd{thin}$.
For DCE 182, although the N$_2$H$^+$ (1--0) emission is very weak and the opacity is not constrained, its emission is most likely optically thin.
The remaining source, DCE 004, is optically thin ($\tau~\sim$ 0.5) in the faintest hyperfine component (110--011, $\tau_{\rm tot}~\times~1/27$). In this source, this component is well separated from the other hyperfine components and a simple Gaussian fitting yields a systemic velocity consistent with the one derived from the HFS fitting.
HCO$^+$ (3--2) and HCN (3--2) are used as optically thick tracers. 
$v_\textmd{thick}$ corresponds to the peak velocity and we assume a 2$\sigma$ uncertainty of 0.14 km s$^{-1}$, which is equal to half of the channel width.
The asymmetry parameters calculated from HCO$^+$ (3--2) and HCN (3--2) are listed in Table 6.

\begin{deluxetable}{cccccc}
\tablewidth{0pt}
\tabletypesize{\tiny}
\tablecaption{Infall signatures} 
\tablehead{ 
\colhead{}    & \multicolumn{3}{c}{$\delta v$} & \colhead{$V_{\rm in}$} \\
\cmidrule(r){2-4}
\colhead{Source}	& \colhead{HCO$^+$}	&\colhead{HCN}		&\colhead{consistency} &\colhead{[km s$^{-1}$]}
}
\startdata
DCE065  &-0.06 (0.15)   & -0.34 (0.16)  & YES	& 0.73 (0.17)\\
DCE004  &-0.78 (0.24)   & -1.23 (0.25)  & YES	& 0.14 (0.02)\\
DCE064  &-0.80 (0.20)   & -0.37 (0.20)  & YES	& -0.62 (0.15)\\
DCE031  &0.53 (0.14)    & 0.63 (0.14)   & YES	& -\\
DCE001  &-0.36 (0.12)   & -0.57 (0.12)  & YES	& 0.07 (0.03)\\
DCE181  &0.43 (0.28)    & -     	& -	& -\\
DCE081  &0.82 (0.18)    & -0.21 (0.18)  & NO	& -0.13 (0.03)\\
DCE025  &1.32 (0.20)    & 1.43 (0.20)   & YES	& -0.08 (0.04)\\
DCE038  &-0.39 (0.18)   & -     	& -	& 0.41 (0.18)\\
DCE185  &1.31 (0.31)    & -1.00 (0.31)  & NO	& -\\
DCE109  &0.16 (0.16)    & -0.13 (0.16)  & NO	& -1.04 (0.15)\\
DCE092  &-0.39 (0.11)   & -0.59 (0.11)  & YES	& 0.58 (0.06)\\
DCE182  &-0.58 (0.23)   & -     	& -	& 0.59 (0.29)\\
DCE107  &0.36 (0.13)    & 0.15 (0.12)   & YES	& -0.46 (0.08)\\
DCE063  &-0.03 (0.11)   & 0.20 (0.11)   & NO	& 0.72 (0.22)\\
DCE090  &-0.08 (0.25)   & -0.54 (0.26)  & YES	& 0.26 (0.05)
\enddata
\tablecomments{This table shows the infall parameters derived for each source. Col.\ (1): Source name. Col.\ (2) and (3): Asymmetry parameters derived from HCO$^+$ (3--2) and HCN (3--2), respectively. Col.\ (4): Consistency label: YES if the asymmetry parameters from HCO$^+$ (3--2) (Col.\ 2) and HCN (3--2) (Col.\ 3) have the same mathematical sign. Col.\ (5): Infall velocity obtained from two-layer model fitting.}
\tablecomments{The values in parenthesis represent 1$\sigma$ uncertainties.}
\end{deluxetable}

We use $|\delta v| \ge 3 \sigma_{\delta v}$ (S/N $\ge$ 3 ) to define the significant detections of blueshifts (infall motion) or redshifts (outward motion).
These sources are separated into three groups, which are 1) $\delta v \le -3\sigma_{\delta v}$: clear indication of infall, 2) $\delta v \ge 3\sigma_{\delta v}$: redshifted optically thick line, and 3) $3\sigma_{\delta v} > \delta v > -3\sigma_{\delta v}$: no clear indication of systematic motion.
With the numbers of sources in the three groups, we calculated the blue excess defined by \citet{ma97},
\begin{equation}
\textmd{blue~excess} = \frac{N_\textmd{blue}-N_\textmd{red}}{N_\textmd{total}},
\end{equation}
to study the statistical properties of our LLO sample.
The blue excess is 0.07 (excluding DCE 185, see below) and 0.15 for HCO$^+$ (3--2) and HCN (3--2), respectively.  
These results are not very different from that the results of previous works, {\it i.e.} 0.25 from \citet{ma97}, 0.28 from \citet{gr00} and 0.05 from \citet{em09}.
We also use the binomial test to determine if the number of blueshifted sources is significant \citep{ry13}.
The survival functions are 0.50 for HCO$^+$ (3--2) and 0.34 for HCN (3--2), suggesting that 
the number of blueshifted detections is not significant in our sample for both HCO$^+$ (3--2) and HCN (3--2).

\begin{figure*}
\includegraphics[scale=.52]{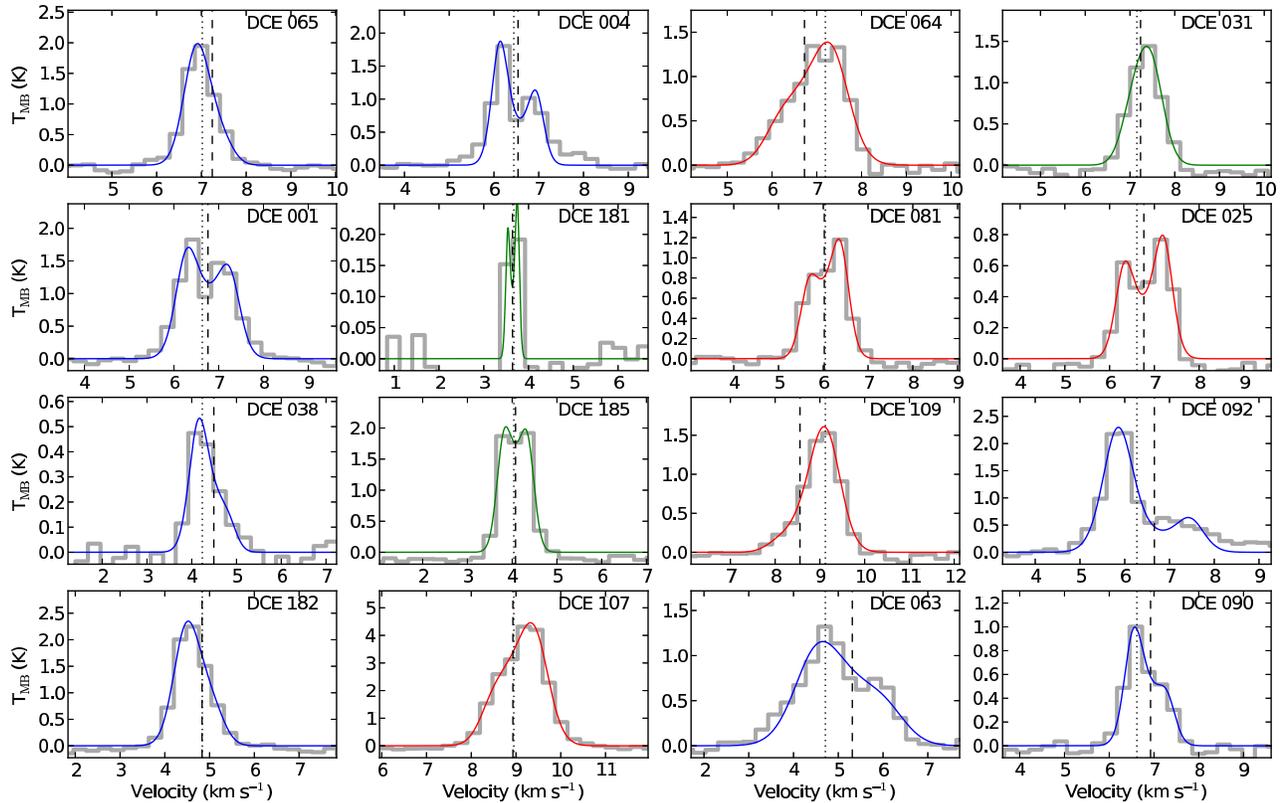}
\caption{Two-layer radiative transfer model fits to the HCO$^+$ (3--2) spectra for the 16 targets. 
The fitting results are shown in blue line for $V_{\rm in}$ $>$ 0, red line for $V_{\rm in}$ $<$ 0 and green for unconstrained $V_{\rm in}$.
The vertical dashed line indicates the $V_{\rm LSR}$ from the two-layer model fitting and the vertical dotted line indicates the $V_{\rm LSR}$ from the N$_2$H$^+$ hyperfine fitting.}
\end{figure*}

\subsection{Two-layer model fitting}
We fitted the HCO$^+$ (3--2) spectra with the two-layer model \citep{my96} to derive the infall velocities of our targets.
We used a simple version defined by \citet{le01}
\begin{equation}
T_{MB} = J(T_r)(1-e^{-\tau_r}) e^{-\tau_f} + J(T_{\rm bg})(e^{-\tau_f-\tau_r} - e^{-\tau_f}),
\end{equation}
where $T_{\rm bg}$ is the cosmic background temperature (2.73 K), $T_r$ is the excitation temperature of the rear layer, and $\tau_f$ and $\tau_r$ are the optical depths of the front and rear layers, respectively. The optical depths are expressed as $\tau_f=\tau_0 \exp[-(v-V_f)^2/2\sigma^2]$ and $\tau_r=\tau_0 \exp[-(v+V_r)^2/2\sigma^2]$, where $\tau_0$ is the peak optical depth, $V_f$ and $V_r$ are the systemic velocities of the front and rear layers, respectively, and $\sigma$ is the velocity dispersion. Considering the relative velocity of the two layers as the infall velocity ({\it i.e.} $V_f = V_r = V_{\rm in}/2$), we fitted the spectra with eq.\ (6) to obtain the infall velocities (Figure 2).
A positive infall velocity represents an inward motion in the core while a negative one indicates an outward motion.
The spectral resolution (0.28 km s$^{-1}$) is sometimes insufficient to resolve the double-peak features and/or asymmetry structures.
We therefore only list the fitting results with infall or outward velocities above the uncertainties (Table 6).
As a result, the infall velocities are positive in 8 sources, negative in 5 sources and unconstrained in 3 sources.

\section{DISCUSSION}
\subsection{The physical conditions from Non-LTE RADEX analysis}

Table 4 shows the physical parameters derived with RADEX. 
The kinetic temperature ($T_{\rm kin}$), N$_2$H$^+$ column density (N(N$_2$H$^+$)), H$_2$ density ($n_{\rm H_2}$) and source sizes of five cores are constrained in the RADEX analysis.
Eight sources have no constraints in their kinetic temperatures and H$_2$ densities;
the data are unable to discriminate models between a wide variety.
Additional spectral lines or higher angular resolution data would be required to discriminate the models for these sources.
Data with higher sensitivity could also break the degeneracy (see Appendix).
The variances of $T_{\rm kin}$ and $n_{\rm H_2}$ from the fitting models are always anti-correlated; the fitting-models can either have high $T_{\rm kin}$ with low $n_{\rm H_2}$ or low $T_{\rm kin}$ with high $n_{\rm H_2}$.
With the currently available data, we discuss the physical properties with the best-fit models from both the 5 constrained sources and the 8 unconstrained sources.
Our LLO targets have $T_{\rm kin}$ in a range between 5.0 K and 33 K, and a median of 8.0 K. (The corresponding thermal line widths are 0.09, 0.24 and 0.12 km s$^{-1}$ for N$_2$H$^+$.) The H$_{\rm 2}$ densities are populated in a range between 10$^5$ cm$^{-3}$ and 2$\times$10$^{6}$ cm$^{-3}$ with a median of 4$\times$10$^{5}$ cm$^{-3}$
(excluding DCE 001 and 185 which have uniquely high $n_{\rm H_2}$ $=$ 2$\times$10$^{7}$ cm$^{-3}$ in our sample, see \S 4.1). 
Comparing with the Class 0 sample ($T_{\rm kin}$ $=$ 9 -- 15 K and $n_{\rm H_2}$ $=$ 5$\times$10$^{6}$ -- 10$^{7}$ cm$^{-3}$) from \citet{em09}, LLOs have relatively lower $n_{\rm H_2}$ and a wider $T_{\rm kin}$ distribution;
the standard deviation of $T_{\rm kin}$ for our LLO sample is $\sim$ 9.3 which is much larger than that of the Class 0 sample, $\sim$ 1.7.
The derived H$_2$ densities of LLOs also spread in a larger range ($\sim$1.3 magnitude) than the Class 0 objects even after excluding DCE 001 and 185.
We think this result comes from a wide variety of natures of our LLO targets which include some Class I objects and are located in different molecular clouds or isolated cores while the Class 0 sources in \citet{em09} are all in the Perseus molecular cloud.
Note that \citet{em09} derived $n_{\rm H_2}$ using N$_2$D$^+$ (2--1) and (3--2) lines, and the kinetic temperatures were derived by \citet{ha03} and \citet{ji99} with NH$_3$ lines.

\begin{figure*}
\includegraphics[scale=.55]{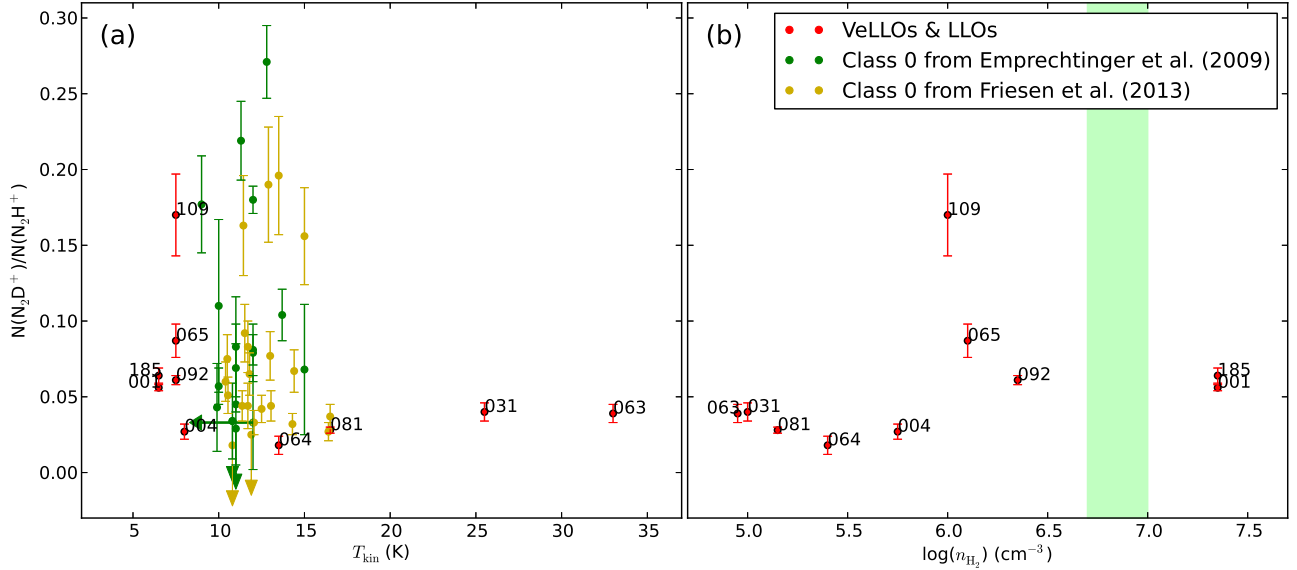}
\caption{Comparison of N$_2$D$^+$/N$_2$H$^+$ column density ratio with $T_{\rm kin}$ and $n_{\rm H_2}$.
The Class 0 data are from \citet{em09} (green) and \citet{fr13} (yellow).
The error bar indicates 1$\sigma$ uncertainty in our data (red). The green area represents the range of H$_2$ densities of the Class 0 objects from
\citet{em09}.
}
\end{figure*}

\begin{figure*}
\includegraphics[scale=.60]{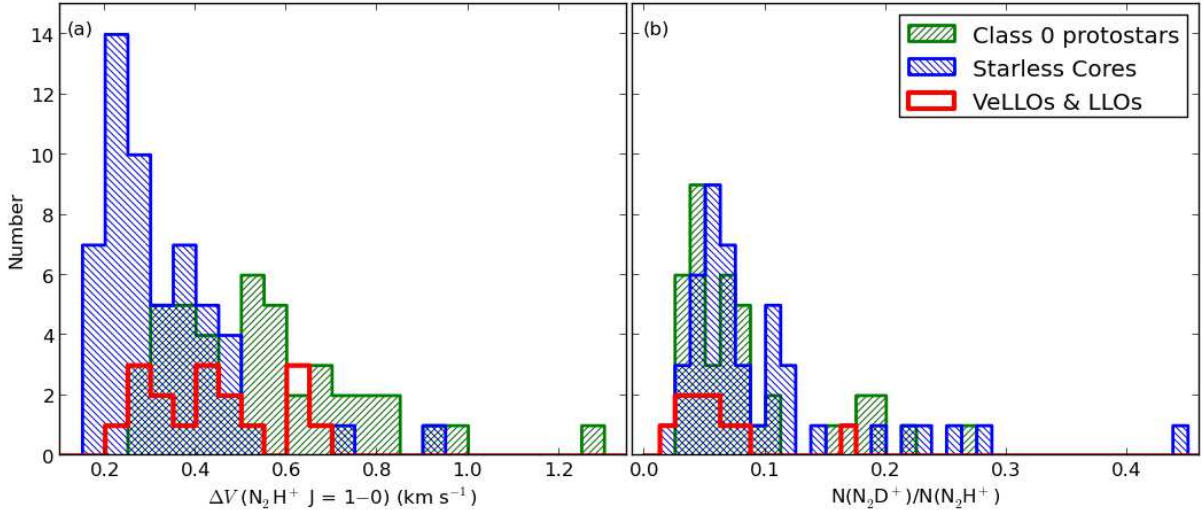}
\caption{Populations of line width (a) and N$_2$D$^+$/N$_2$H$^+$ column density ratio (b) for Class 0 protostars (green), Starless cores (blue) and LLOs (red). The Class 0 protostar and starless core data are from \citet{em09}, \citet{fr13} and \citet{cr05b}.
}
\end{figure*}

A caveat in the H$_2$ density comparison is that the beam size of our data is much larger than that from \citet{em09}.
The larger beam could make the derived H$_2$ density lower if it is diluted by a large area.
To fairly compare the H$_{\rm 2}$ densities, we scale our derived $n_{\rm H_2}$ to the observing beam size of N$_2$D$^+$ (2--1) (16.3\arcsec) from \citet{em09}.
Assuming a density profile, $\rho$(r) $\propto$ r$^{-2}$, we calculate an average density, $\rho_{\rm ave}$($<$r) $\propto$ r$^{-2}$, within a radius r.
We can therefore obtain an expected H$_2$ density within 16.3\arcsec~area based on our derived H$_2$ density and source size.
The scaled H$_2$ density is populated in a range between 7$\times$10$^{4}$ cm$^{-3}$ and 3$\times$10$^{7}$ cm$^{-3}$ with a median of 2$\times$10$^{6}$ cm$^{-3}$.
Our derived H$_2$ densities therefore have a wider distribution compared with the Class 0 sample.
Excluding DCE 001 and 185 (see \S 4.1), 3 LLOs have H$_2$ density higher than Class 0 sources and 8 have lower H$_2$ density.
Therefore, LLOs are likely to have relatively low H$_2$ densities compared with Class 0 sources after correcting for the different beam sizes.

\begin{figure*}
\includegraphics[scale=.5]{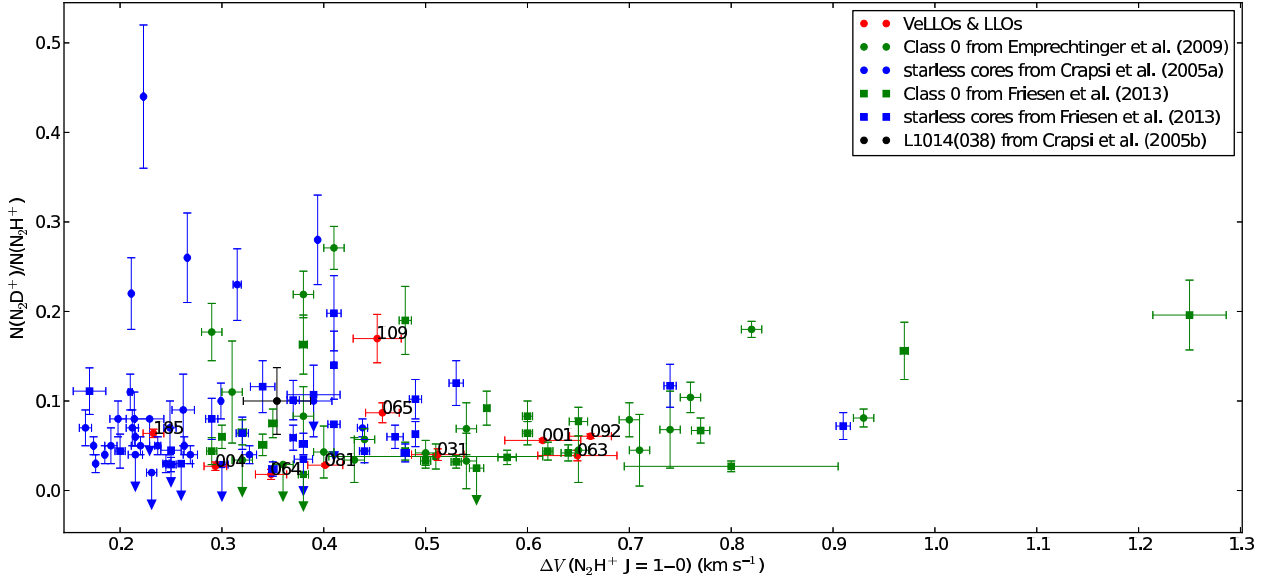}
\caption{Line widths of N$_2$H$^+$ (1--0) versus N(N$_2$D$^+$)/N(N$_2$H$^+$) ratios. The colors represent the different type of sources as Fig.\ 2.
The error bars indicate 1$\sigma$ uncertainties in our data (red).
}
\end{figure*}

\subsection{N(N$_2$D$^+$)/N(N$_2$H$^+$) ratio as an age indicator}
To investigate the relation between chemistry and physical conditions in these cores,
we compare the N(N$_2$D$^+$)/N(N$_2$H$^+$) ratio with the derived kinetic temperature (or gas temperature) and H$_2$ density (Figure 3).
Figure 3a includes the Class 0 objects from \citet{em09} and \citet{fr13}.
The kinetic temperatures are derived using N$_2$H$^+$ J $=$ 3--2/J $=$ 1--0 in our study, and using NH$_3$ (1, 1)/ (2, 2) in \citet{em09} and \citet{fr13}.
The sources with the highest N(N$_2$D$^+$)/N(N$_2$H$^+$) ratios ($>$ 0.05) all have low kinetic temperatures ($<$ 10 K).
We use Pearson's $r$ correlation test and Spearmans's $\rho$ rank correlation test \citep{co99} to evaluate the significances of correlations between N(N$_2$D$^+$)/N(N$_2$H$^+$) ratios and kinetic temperatures for our LLO sample.
The Pearson correlation coefficient, $r \sim$ -0.38 with a significance of 0.28, provides no evidence of (linear) correlation between $T_{\rm kin}$ and N(N$_2$D$^+$)/N(N$_2$H$^+$) ratio.
However, the Spearman rank correlation coefficient, $\rho \sim$ -0.61 with a significance of 0.06 (the probability for this sample to be an uncorrelated system), implies an existence of anti-correlation.
This features are predicted by chemical models; the N(N$_2$D$^+$)/N(N$_2$H$^+$) ratio is expected to increase in cold regions.
However, we did not find any significant correlation between the N(N$_2$D$^+$)/N(N$_2$H$^+$) ratio and $n_{\rm H_2}$ (Figure 3b).

To study the evolutionary states of LLOs, we compare the chemical evolutionary indicator, N(N$_2$D$^+$)/N(N$_2$H$^+$) ratio, and the dynamical evolutionary indicator, N$_2$H$^+$ (1--0) line width, with that of starless cores and Class 0 sources from \citet{cr05b}, \citet{em09} and \citet{fr13}.
Figures 4a and 4b show the line width and N(N$_2$D$^+$)/N(N$_2$H$^+$) ratio populations, respectively.
The line widths of Class 0 sources are relatively larger than that of starless cores, and most of our LLOs are located in the region where Class 0 sources and starless cores overlap (Figure 4a).
This result implies that LLOs are likely young Class 0 sources whereas there is no significant difference in their N(N$_2$D$^+$)/N(N$_2$H$^+$) ratio (Figure 5).
The N(N$_2$D$^+$)/N(N$_2$H$^+$) ratios (median$=$0.05) of our LLOs are similar to that of starless cores and Class 0 sources, although only 10 sources have N$_2$D$^+$ (3--2) detections. (Though N$_2$D$^+$ J = 3 $\rightarrow$ 2 is detected in DCE 107, the two components are unresolvable, which prevents us from deriving the N$_2$D$^+$ column density for each component.)

\subsection{Infall indicators}

Figure 6 shows the comparison of the asymmetry parameters calculated from HCO$^+$ (3--2) and HCN (3--2). 
The asymmetry parameters obtained from HCO$^+$ (3--2) and HCN (3--2) are quite consistent with each other except for DCE 185;
HCO$^+$ (3--2) $\delta v$ and HCN (3--2) $\delta v$ show a positive correlation.
For the 13 targets with both HCO$^+$ (3--2) and HCN (3--2) detections, 9 targets have the same mathematical sign (Table 6).
The remaining 4 sources are DCE 081, 185, 109 and 063.
The asymmetry parameters of DCE 109 and 063 have S/N ratios less than 2$\sigma$ for both HCO$^+$ (3--2) and HCN (3--2), which implies that no clear systematic motions are occurring (Figure 6).
The HCO$^+$ (3--2) spectrum of DCE 185 shows a double-peak profile and the peak values are very similar. 
If we used the alternative peak, the asymmetry parameter would become -1.12, which would be consistent to that from HCN (3--2).
However, it is still unclear if DCE 185 has a clear infall or redshifted signature based on the low spectral resolution data.
The DCE 081's HCN (3--2) $\delta v$ indicates no clear systematic motion ($\sim 1.2 \sigma$) and its HCN (3--2) detection is quite weak (S/N $\sim$ 4.2).
Since its HCO$^+$ (3--2) spectrum shows clear asymmetry profile (see below), we suggest that its asymmetry parameter from HCO$^+$ (3--2) is much more reliable than that from HCN (3--2) for DCE 081.

Using the two-layer model, the infall velocities of 13 targets are derived using the HCO$^+$ (3--2) spectra.
We compared the derived infall velocities with previous works.
\citet{be02} have derived an infall velocity field in DCE 001 (IRAM 04191). The infall velocity is decreasing from the center and becomes uniform ($V_{\rm in} \sim$ 0.10 $\pm$ 0.05 km s$^{-1}$) between $\sim$2000--3000 AU and $\sim$10000--12000 AU.
Our derived infall velocity, 0.07 $\pm$ 0.03 km s$^{-1}$ , is consistent with this infall velocity field in the outer region.
For DCE 004 (L1521F), our derived infall velocity 0.14 $\pm$ 0.02 is slightly smaller than the infall velocity 0.2--0.3 km s$^{-1}$ (2000 to 3000 AU) from \citet{on99}, but much larger than that from \citet{le04}, 0.014 km s$^{-1}$ and 0.045 km s$^{-1}$ from CS (2--1) and (3--2) spectra, respectively.
\citet{le04} concluded that the CS (2--1) and (3--2) may trace different regions, which results in a difference between the derived infall velocities.
HCO$^+$ (3--2) may trace a more excited region since it has a higher energy level.
In addition, the difference of optical depth and depletion between molecules may also explain the discrepancy between infall velocity measurements.

We also compared the infall velocities with the free-fall velocities. The free-fall velocities, $v_{\rm ff}= (GM/R)^{1/2}$ \citep{ke04} where $M$ and $R$ are the
envelope mass and radius, respectively,
are derived from the RADEX best-fit models assuming the central star has a negligible mass.
All sources have infall velocities less than free-fall velocities except for DCE 038, 063 and 090. 
The infall velocities of these three sources are not significantly larger than their free-free fall velocities (within 3$\sigma$ uncertainty of $V_{\rm in}$). Thus, none of our targets has infall velocity significantly larger than its free-fall velocity.

All the sources with positive $V_{\rm in}$ have negative HCO$^{+}$ (3--2) $\delta v$ and vice versa except for DCE 064.
This result suggests that the infall or red shifted (outward) signatures are quite reliable in the HCO$^{+}$ infall analysis.
For DCE 064, the asymmetry spectrum with blueshifted wing suggesting an expanding motion is in contrast to the negative $\delta v$ derived from both HCO$^+$ (3--2) and HCN (3--2).
The spectral resolution of our current data is insufficient to identify if DCE 064 has an infalling or expanding envelope.
We find that the HCO$^+$ spectrum of DCE 064 has a double-peak profile, and the peak values are similar as DCE 185.
The HCO$^+$ $\delta v$ will be $\sim$ 0.8 (expanding) if we use the second peak whereas the $\delta v$ derived from the first peak is -0.78.
In addition, the asymmetry parameter of HCN (3--2), $\delta v$ = -0.37, is not significant (S/N $\sim$ 1.8).
Therefore, the blue shifted wing hints that outward motion may occur in DCE 064.

It is noteworthy that
the HCO$^+$ systemic velocities from the two-layer model fitting are occasionally shifted from the systemic velocities obtained from N$_2$H$^+$ (Figure 2).
These results suggest that HCO$^+$ and N$_2$H$^+$ trace different regions in the cores, and hint that the derived asymmetry parameters could be false.
Observations of optically thin isotopologue H$^{13}$CO$^+$ or HC$^{18}$O$^{+}$ are required to correct the infall measurements by providing an unbiased $V_{\rm LSR}$.
HCO$^+$ can also trace the protostellar outflows which could contaminate our determinations of infall motions.
Besides, more complicated dynamic structures such as molecular outflows or core rotation could affect our results.
In addition, \citet{sm12} found that the blue asymmetry parameters could be frequently undetected toward collapsing cores located in filaments.
Based on radiative transfer models, \citet{sm12} found that geometry parameters such as inclination angle will significantly affect the profile of optically thick tracers like CS (2--1) and HCN (1--0).
In order to find a reliable infall indicator, \citet{ch14} modeled the line profiles of HCO$^+$ and HCN among transitions from J = 1 $\rightarrow$ 0 to J = 5 $\rightarrow$ 4.
They found that J = 4 $\rightarrow$ 3 has the best chance to probe infall motions, but the J = 3 $\rightarrow$ 2 and J = 5 $\rightarrow$ 4 transitions of both HCN and HCO$^+$ are also good indicators.

As a result, 8 sources have positive infall velocities (infall motions) and 5 sources have negative infall velocities (outward motions).
Out of the 8 infall candidates, 3 sources (DCE 004, 001, and 092) have significant blue asymmetry (S/N $\ge$ 3) from both HCO$^+$ (3--2) and HCN (3--2).
These 3 sources have very low kinetic temperatures (6.5 K $<$ $T_{\rm kin}$ $<$ 8.0 K) and bolometric temperatures (20 K $<$ $T_{\rm bol}$ $<$ 27 K).
Four sources (DCE 065, 038, 182, and 090) have a blue asymmetry parameter in either HCO$^+$ (3--2) or HCN (3--2) in the range 2--3$\sigma$, which suggests that infall motions may be occurring in these candidates too.
DCE 065 and 038 also have low kinetic temperatures (7.5 and 9.0 K) and bolometric temperatures (29 and 66 K).
DCE 182 and 090 have relatively high bolometric temperatures (105 and 114K, {\it i.e.} young Class I sources), and DCE 090 has a high kinetic temperature of 30.5 K ($T_{\rm kin}$  of DCE 182 cannot be solved due to lack of N$_2$H$^+$ (3--2)). 
Our results suggest that the infall motions are likely to occur at an early evolutionary stage but can still appear at a later stage (Class I), which is consistent with \citet{gr00}.

\begin{figure}
\includegraphics[scale=.48]{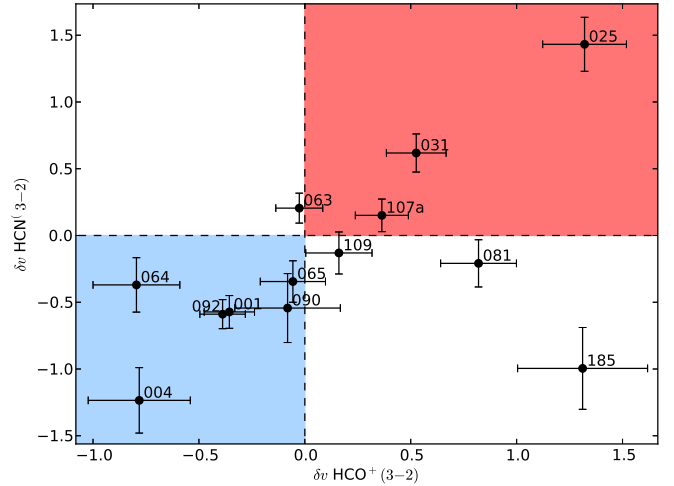}
\caption{
Asymmetry parameters derived from HCO$^+$ (3--2) versus that from HCN (3--2).
The blue and red colors indicate the area for infall motion and outward motion, respectively.
The error bars indicate 1$\sigma$ uncertainties.
}
\end{figure}

\section{CONCLUSION}
We observed N$_2$H$^+$ (1--0), N$_2$H$^+$ (3--2), N$_2$D$^+$ (3--2), HCO$^+$ (3--2) and HCN (3--2) toward 16 LLOs using the KP 12m telescope and SMT in ARO.
We derived the kinetic temperatures, N$_2$H$^+$ column densities, H$_2$ densities and core sizes by fitting spectra with the non-LTE code RADEX.
We obtain these physical parameters for 13 LLOs from N$_2$H$^+$ (1--0) and (3--2) observations.
We derived the N(N$_2$D$^+$)/N(N$_2$H$^+$) ratio which traces the evolutionary state of starless cores and protostellar cores.
To further probe the infall motions,
we used HCO$^+$ (3--2) and HCN (3--2) as optically thick tracers and N$_2$H$^+$ (1--0) as optically thin tracer.
Our conclusions are the following:

\begin{enumerate}
\item  
LLOs tend to have relatively low H$_2$ densities compared with the Class 0 objects in \citet{em09}.
Only DCE 001 and 185 have higher H$_2$ densities than Class 0 objects;
however, their densities are most likely overestimated due to the beam size difference in N$_2$H$^+$ (1--0) and (3--2) observations.
\item 
For $T_{\rm kin} < 10$ K, LLOs have N$_2$D$^+$/N$_2$H$^+$ column density ratios between 0.03--0.17 similar to the Class 0 objects observed by \citet{em09} and \citet{fr13}, and for $T_{\rm kin} > 10$ K, LLOs have consistently low N$_2$D$^+$/N$_2$H$^+$ column density ratios $\sim$ 0.04.
The Spearman rank correlation test implies an anti-correlation between the N$_2$D$^+$/N$_2$H$^+$ column densities and the kinetic temperatures in LLOs with a significance of 0.06.
\item 
We find that the N$_2$D$^+$/N$_2$H$^+$ column density ratios of LLOs are similar to that of starless cores and Class 0 objects,
but the line widths of LLOs are mostly populated in the starless cores and Class 0 objects overlapping region.
We suggest that the molecular line width could be a better age indicator than N(N$_2$D$^+$)/N(N$_2$H$^+$) ratio at the early stage and
LLOs are likely young Class 0 protostellar sources.
\item 
We identify eight infall candidates by fitting the HCO$^+$ (3--2) spectra with a two-layer model. 
Seven sources are supported by their asymmetry parameters with S/N $>$ 2 in either or both HCO$^+$ (3--2) or HCN (3--2).
Our result suggests that infall motions tend to occur at a very early evolutionary stage but can still occur in young Class I sources (DCE 090 and 182).

 \end{enumerate}

\acknowledgments
T.H.H. and S.P.L. acknowledge support from the Ministry of Science and
Technology (MOST) of Taiwan with Grants NSC 101-2119-M-007-004
and MOST  102-2119-M-007-004-MY3.
T.H.H thanks to MOST for granting him the PhD exchanging
student award (MOST 103-2917-I-007-005) to visit MPIfR from Feb 2014 to Jan 2015, and is also indebted to Prof. Dr. Karl M. Menten
for his visiting in MPIfR.


\appendix
We evaluated the feasibility of using N$_2$H$^+$ (1--0) and (3--2) data to well constrain the physical parameters with RADEX.
The method is to find the relation between the uncertainties of $T_{\rm kin}$ and $n_{\rm H_2}$ and a set of synthetic spectra with different rms noise levels.
We first construct a reference model using RADEX and compute the synthetic N$_2$H$^+$ (1--0) and (3--2) spectra for this reference model. Then we added artificial noise to the spectra and derived the excitation temperatures and optical depths using the HFS fitting method. The data with lower rms noise levels would have lower uncertainties in their excitation temperatures and optical depths.
We made grid of models with varied kinetic temperatures, H$_2$ densities, and N$_2$H$^+$ column densities. Then we compute the $\chi^2$ for the reference model and each model in the grid based on the excitation temperatures and optical depths.

\setcounter{figure}{0}
\renewcommand{\thefigure}{A\arabic{figure}}

\begin{figure}
\includegraphics[scale=.43]{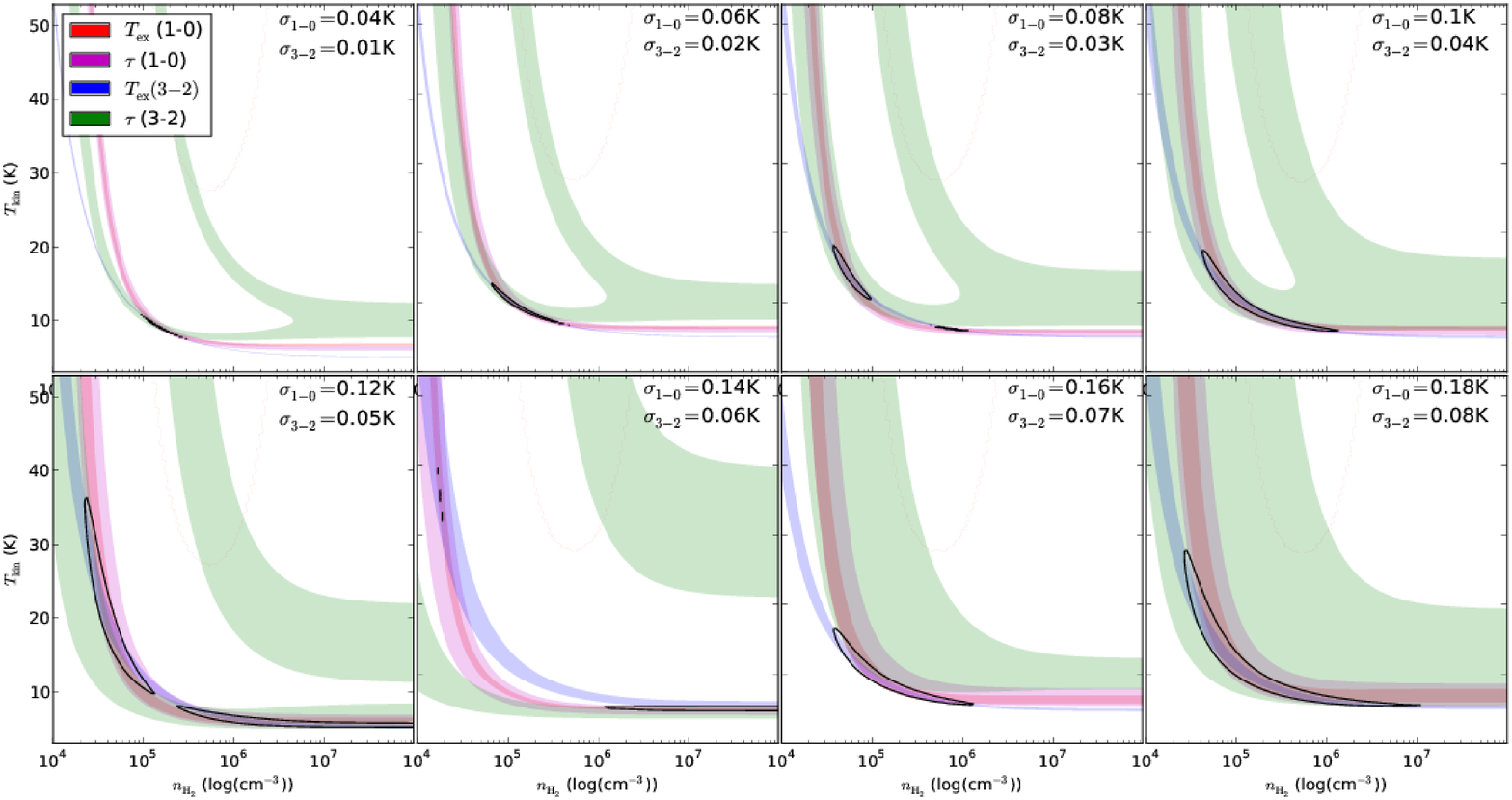}
\caption{The slice in the model grid for $T_{\rm kin}$ versus $n_{\rm H_2}$. The black line indicates the confidence interval which fits to the synthetic spectra. The color area indicates the models fit to the observing parameters, $T_{\rm ex}$ and $\tau$, within their errors.}
\end{figure}

Here we describe the process and results in detail.
Using RADEX without hyperfine structure, we derive $T_{\rm ex}$ and $\tau$ for both N$_2$H$^+$ (1--0) and (3--2)
from a reference model set, T$_{\rm kin} =$ 8.75 K, log($n_{\rm H_2}) =$ 5.225, N(N$_2$H$^+$) $=$ 1.1$\times$ 10$^{13}$ cm$^{-2}$ and core size $=$ 75\arcsec.
We constructed the synthetic N$_2$H$^+$ (1--0) and (3--2) spectra with
\begin{equation}
T_{\rm MB} (v) = \Psi (J(T_{\rm ex}) - J(T_{\rm bg}))(1-exp(-\tau(v))),
\end{equation}
considering the beam sizes of 67.5\arcsec~ for N$_2$H$^+$ (1--0) and 27\arcsec~ for N$_2$H$^+$ (3--2).
We added noise into the synthetic spectra. 
The noise levels are from 0.04 to 0.16 K (at a spectral resolution of 0.078 km s$^{-1}$) with a step of 0.02 K for N$_2$H$^+$ (1--0) and from 0.01 to 0.07 K (at a spectral resolution 0.27 km s$^{-1}$) with a step of 0.01 K for N$_2$H$^+$ (3--2); in total, seven sets of synthetic spectra are constructed. (The rms noise level of our observational data are $\sim$0.1 K and $\sim$0.04 K for these two transitions.)
We obtain $T_{\rm ex}$ and $\tau$ using HFS fits to the synthetic spectra.
As expected, the errors of $T_{\rm ex}$ and $\tau$ are correlated to the artificial rms noise levels.
We then construct a grid of models with varied kinetic temperatures, H$_2$ densities, and N$_2$H$^+$ column densities.
We calculate the $\Delta \chi^2$ using the $T_{\rm ex}$ and $\tau$ from the synthetic spectra of the reference model and that from each model in the grid, {\it i.e.} $\Delta \chi^2 = (\frac{T_{\rm ex, 1-0}-T_{\rm ex, 1-0, model}}{\sigma_{T_{\rm ex, 1-0}}})^2 + (\frac{\tau_{1-0}-\tau_{\rm 1-0, model}}{\sigma_{\tau_{1-0}}})^2 + (\frac{T_{\rm ex, 3-2}-T_{\rm ex, 3-2, model}}{\sigma_{T_{\rm ex, 3-2}}})^2 + (\frac{\tau_{3-2}-\tau_{\rm 3-2, model}}{\sigma_{\tau_{3-2}}})^2$.
Figure A1 shows the 68.3\% confidence interval in the slices of $T_{\rm kin}$ vs. $n_{\rm H_2}$ space.
The color regions represent the models fitting the observing parameters within the errors.
The regions become broader when the uncertainties become larger with larger rms noise levels.
With a fixed $T_{\rm ex}$ or $\tau$, the fitting models are always populated from high $T_{\rm kin}$ with low $n_{\rm H_2}$ to low $T_{\rm kin}$ with high $n_{\rm H_2}$.
This phenomenon explains that the models with high $T_{\rm kin}$ low $n_{\rm H_2}$, and low $T_{\rm kin}$ high $n_{\rm H_2}$ are sometimes discriminated with difficulty.
As a result, we provided the reference values of noise for deriving the physical parameters.
For example, the data set with best sensitivity, $\sigma_{1-0} \sim$ 0.04 K and $\sigma_{3-2} \sim$ 0.01 K, can be used to constrain the $T_{\rm kin}$ in a range of $\sim$ 7-11 K and the $n_{\rm H_2}$ in a range of $\sim$ 6.3 $\times$ 10$^{4}$ - 1.6 $\times$ 10$^{5}$ cm$^{-3}$ (Figure A1). 
Note that if we fit the spectra with HSF directly, the degeneracy can be marginally reduced (see \S 4.1). 
Our analysis provides the reference sensitivity which is required for constraining the physical conditions with RADEX. 
However, the analysis does not guarantee that the sensitivity allows to achieve the constraint as our example. 
If the best-fit model is located in a different region, the confidence interval is expected to be different.

\begin{figure}
\includegraphics[scale=.43]{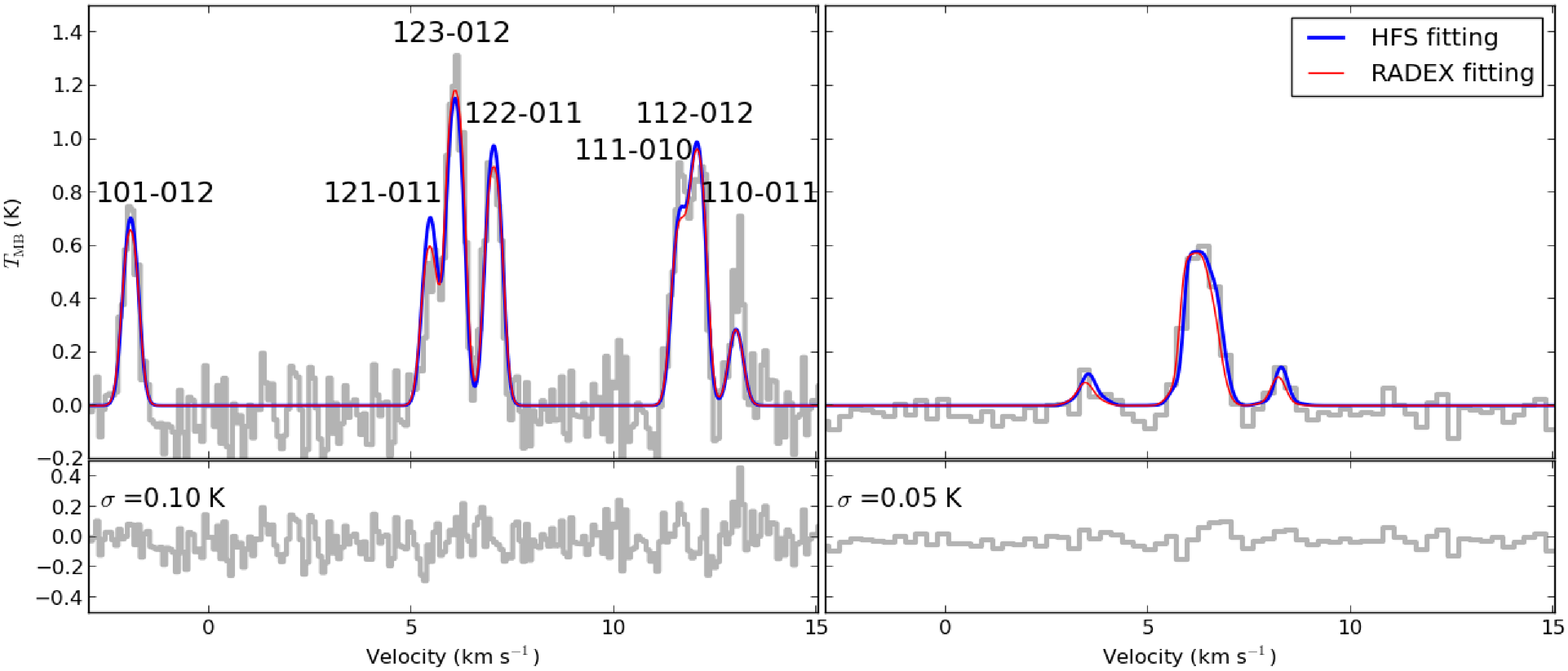}
\caption{Comparison of HFS fitting (blue) and RADEX fitting (red) for DCE 081. The left and right panels show the N$_2$H$^+$ (1--0) and (3--2) spectra of DCE 081, respectively. The bottom panels show the residuals of the RADEX fit.}
\end{figure}

\end{document}